\documentclass[aps,prb,twocolumn,superscriptaddress,floatfix]{revtex4-2}
\usepackage{blindtext}
\usepackage{graphicx}
\usepackage{bm}
\usepackage{hyperref}
\usepackage{amsmath}
\usepackage{cancel}

\usepackage{xcolor}

\graphicspath{{Figs/}}

\usepackage{cleveref}

\usepackage{hyperref}
\hypersetup{%
	pdftitle={cavity}, %
	pdfauthor={Zhiyuan Sun},%
	pdfpagemode={UseNone},%
	pdfstartview={FitH},%
	breaklinks=true,%
	citecolor=blue,%
	colorlinks=true,%
	linkcolor=blue,%
	urlcolor=blue
}

\newcommand{\unit}[1]{\,\mathrm{#1}} 
\newcommand{\equa}[1]{Eq.~\eqref{#1}} 
\newcommand{\fig}[1]{Fig.~\ref{#1}}

\newcommand{\sect}[1]{Sec.~\ref{#1}}
\newcommand{\append}[1]{Appendix~\ref{#1}}
\usepackage{ulem}

\newcommand{\rom}[1]{\uppercase\expandafter{\romannumeral #1\relax}}

\renewcommand{\vec}{\mathbf}

\usepackage{xcolor}
\definecolor{darkred}{RGB}{139,0,0}

\usepackage[T1]{fontenc}
\usepackage{times}{\Large}

\begin{document}
\title{Floquet polaritons in optically driven materials}

\author{Tsan Huang}
\author{Teng Xiao}
\affiliation{State Key Laboratory of Low-Dimensional Quantum Physics and Department of Physics, Tsinghua University, Beijing 100084, P. R. China}

\author{Jiahua Duan}
\affiliation{State Key Laboratory of Chips and Systems for Advanced Light Field Display, MOE Key Lab of Advanced Optoelectronic Quantum Architecture And Measurement, School of Physics, Beijing Institute of Technology, Beijing, China}

\author{Haoliang Qian}
\email{haoliangqian@zju.edu.cn}
\affiliation{State Key Laboratory of Extreme Photonics and Instrumentation, College of Information Science and Electronic Engineering, Zhejiang University, Hangzhou, 310027, China.}

\author{Zhiyuan Sun}
\email{zysun@tsinghua.edu.cn}
\affiliation{State Key Laboratory of Low-Dimensional Quantum Physics and Department of Physics, Tsinghua University, Beijing 100084, P. R. China}
\affiliation{Frontier Science Center for Quantum Information, Beijing 100084, P. R. China}

\begin{abstract}
Polaritons are coupled collective modes of light and matter in quantum materials.
In modern pump-probe experiments, a pump light pulse may dramatically alter the properties of the polaritons, rendering them Floquet polaritons that can be detected by a probe pulse.
We present a practical framework to describe Floquet polaritons in terms of the linear and nonlinear optical properties of the material. 
The central quantity that yields the spectra of Floquet polaritons is an effective linear optical susceptibility contributed by the pump through nonlinear optical susceptibilities.
We apply this method to graphene and show that via its third-order optical nonlinearity, infrared pump leads to Floquet plasmon bands.
Notably, near plasmonic band crossings, parametric instability leads to flat bands with unstable modes and exceptional points that closely resemble those of non-Hermitian systems.
As a second example, we show that in hexagonal boron nitride pumped by  mid-infrared laser, the pump induces Floquet phonon polariton bands via  phononic nonlinearity,  which can be detected with either far-field or near-field optical technique.
Finally, in layered superconductors pumped by THz light polarized along the out-of-plane direction, the Josephson-type optical nonlinearity leads to Floquet Josephson plasmons,  which manifest as new peaks in the THz reflectivity of a probe pulse.
\end{abstract}

\maketitle

\section{Introduction}

Polaritons are coupled collective modes between matter degrees of freedom and electromagnetic field~\cite{basov2016polaritons,low2017polaritons, Basov2021PolaritonPanorama, deAbajo2025Roadmap2DPhotonics}. 
In the quantum mechanical language,  they are bosonic excitations as mixtures of matter excitations and photons in materials.
Polaritons have been discovered experimentally in a wide range of systems,
including plasmons~\cite{maier2007plasmonics,
Fei2012GraphenePlasmons,	
Chen2012GraphenePlasmons,
grigorenko2012graphene,
Woessner2015GrapheneBNPlasmons,
Hu2016GraphenePlasmonMolecularFingerprints,
Ni2018FundamentalLimitsGraphenePlasmonics,
Zhao2023HydrodynamicPlasmonsGraphene}, 
phonon polaritons~\cite{huang1951lattice,dai2014tunable,
Caldwell2014SubDiffractionalHBNPolaritons,	
Caldwell2015SurfacePhononPolaritons, Li2015HBNHyperbolicPolaritons, 
Yang2016CoupledPlasmonPhononPolaritons,
nemilentsau2016anisotropic,
Giles2018UltralowLoss, 
Li2018InfraredHyperbolicMetasurface,
Ni2021LongLived, 
Guddala2021TopologicalPhononPolariton,
Kurman2021SpatiotemporalImaging,
Ma2021GhostHyperbolicSurfacePolaritons,
Guo2023WhisperingGalleryHBNPolaritons,
Hu2023GateTunableNegativeRefractionPolaritons,
Duan2023PhotonicMagicAngles,
Sun2024QuaternaryOxidesPolaritons,
Guo2025HyperbolicElectroluminescence}, 
exciton polaritons~\cite{hopfield1958theory, Kasprzak2006PolaritonBEC, deng2010exciton, Carusotto2013QuantumFluids,
Su2017PerovskitePolaritonLasing,
Su2020PerovskitePolaritonCondensation,
song2025room}, etc.
They also exist in systems with nontrivial electronic correlations such as
magnon polaritons~\cite{Huebl2013MagnonPhoton,Tabuchi2014MagnonPhoton} in magnets, 
Josephson plasmons 	and other polaritons in superconductors~\cite{Saveliev2010JosephsonPlasmaWaves,
Laplace2016JosephsonPlasmonics, sun2020collective,nicoletti2022coherent,kaj2023terahertz,zhang2023revealing,
Sellati2023GeneralizedJosephsonPlasmons,Fiore2024InvestigatingJosephsonPlasmons}, and those in excitonic insulators~\cite{Murakami.2020, SunMillis2020BardasisPolaritons, Xuan.2026, shao2026electromagneticresponsesbilayerexcitonic}. 
In addition to being probes of the fundamental physics of the underlying material, polaritons are also promising in technological applications  as nanoscale information carriers and ultrafast optical transistors, and even as carriers and processors of quantum information~\cite{ AlonsoCalafell2019QuantumComputingGraphenePlasmons,
Ghosh2020QuantumComputingPolaritons,
Sun_plasmon.2022,
Calajo2023NonlinearQuantumLogicGraphenePlasmons,
dapolito2026quantumlightnanoimaging}.
So far, experiments on polaritons have largely focused on the weak field regime which probes the linear response of the system, which is in principle  an equilibrium property.
On the other hand, the emergence of ultrafast and ultrastrong laser techniques~\cite{
	Wagner2014UltrafastNanoscaleGraphenePlasmons,
	Ni2016UltrafastGraphenePlasmons, Basov2017PropertiesDemand} has brought the possibility of  exploring non-equilibrium physics with polaritons.
For example, a strong pump-laser pulse may dramatically alter the properties of these polaritons, opening a new dimension for engineering novel polaritons and their applications on ultrafast time scales.
	
Along this line, the most natural question is how the oscillating optical field of the strong pump laser engineers the polaritonic spectra, transforming them into `Floquet polaritons'.
The use of strong light fields to engineer new states of matter, 
Floquet engineering~\cite{oka2009photovoltaic,
Bukov2015FloquetReview,
oka2019floquet,
rudner2020floquetengineershandbook,
rodriguez2021low,
Bao2022LightInducedEmergent,
mori2023floquet,
boschini2024time}, 
has emerged as a powerful tool for controlling electronic band structure~\cite{
oka2009photovoltaic,
Kitagawa2011PhotoInducedQH,	
lindner2011floquet, sentef2015theory,
hubener2017creating,harper2020topology,
rudner2020band,
xiao2025interactioneffectselectronicfloquet} and correlation~\cite{Ponte2017many,kennes2018floquet,kennes2019light,
DelaTorre2021UltrafastControl,
sun2024floquet,
huang2025universalphasetransitionsmatter,
dong2026opticaldetectionmanipulationpseudospin}
in quantum materials.
Recently, exciting progress  has made on engineering electronic band structures in, e.g., topological insulators~\cite{wang2013observation}, black phosphorus~\cite{zhou2023pseudospin} and graphene~\cite{mciver2020light, ito2023build, Wang2026FloquetGapGraphene}. 
However, Floquet engineering of polaritons by light has remained largely unexplored.

In this work, we show a practical theoretical method for predicting Floquet polaritons  in quantum materials under pump laser.
The central method is to derive the effective linear optical susceptibility $\chi_{\text{eff}}$ of the system that is modified by the pump through optical nonlinearities.
The spectra of the  Floquet polaritons  are then obtained by solving the Maxwell's equations with this effective susceptibility.	
Because the latter is constructed in terms of the pump field and the nonlinear optical coefficients of the system, this approach allows unambiguous prediction of the Floquet polaritonic spectra without artificial parameters.

We apply this method to three representative systems driven by pump light.
The first system is graphene with doped carriers which hosts infrared plasmons. We show that through third-order optical nonlinearity, the  pump light transforms the plasmons into Floquet plasmons, which exhibit flat bands with unstable modes and exceptional points similar to non-Hermitian systems~\cite{MiriAlu2019ExceptionalPoints,Ozdemir2019ExceptionalPoints}.
These phenomena can be understood as arising  from parametric instability that renders the system a gain medium.
In the second system, hexagonal boron nitride (hBN) that hosts hyperbolic phonon polaritons~\cite{Dai2019MonolayerhBNPolaritons,Li2021SuspendedMonolayerhBNPolaritons}, we show that pumping it at roughly twice the transverse optical (TO) phonon frequency leads to similar parametric instability and 
non-Hermitian bands for Floquet phonon polaritons, both in hBN monolayers and thin flakes.
In addition, resonant pumping of the TO phonon creates new phonon polariton modes around twice the TO frequency in monolayer hBN. 
Finally, we apply the method to the third system, a layered superconductor pumped with THz light along the out-of-plane direction.
We show that the pump leads to Floquet Josephson plasmons which manifest as new peaks in the THz reflectivity of a probe pulse.
These Floquet polaritons are shown to be experimentally accessible through pump-probe protocols within reach of current experimental capabilities.

Floquet engineering of polaritons has been discussed before in cold atomic systems~\cite{Clark2019InteractingFloquetPolaritons, Johansen2022MultimodeFloquetPolaritonSuperradiance}, photonic crystals~\cite{Galiffi2022TimeVaryingMedia, Ozlu.2025}, and for excitons~\cite{Liew.2018}.	
However, quantum materials driven by pump light provide the most feasible experimental realization and the most promising platform for technological applications of Floquet polaritons, especially in the near field regime.
There the periodic driving of the polaritons is fixed by the nature of light-matter coupling.
Our work goes beyond previous ones in two aspects.
First, by treating the pump effect using nonlinear optical coefficients which are experimentally measurable, our approach offers practical, unambiguous and quantitative  predictions of Floquet polaritons  without artificial driving parameters.
As a result, new Floquet phenomena are found which are also experimentally relevant.
Second, unlike Floquet engineering of electrons, the Floquet polariton problem often lacks a Hamiltonian description.
This is because polaritons, described in the photonic language,  are in principle open and dissipative systems obtained by integrating out the matter degrees of freedom.
Our Green function approach, based on the Keldysh path integral, is the rigorous formalism for this problem of driven open dissipative systems. 
To help readers, we use Green functions based on the equation of motion  in the main text that is the classical approximation to the full Keldysh path integral formalism, and keep the latter in  \append{app:keldysh}.

The rest of the paper is organized as follows.
In \sect{sec:formalism}, we develop the general formalism for Floquet polaritons in optically driven materials.
We first formulate polaritonic propagators from Maxwell's equations and the equilibrium optical susceptibility, and then show how the pump field combines with nonlinear optical susceptibilities to generate an effective linear susceptibility $\chi_{\text{eff}}$ for the probe field.
In \sect{sec:gplasmon}, we apply this framework to doped graphene and derive the Floquet plasmon bands induced by the pump via the third-order optical nonlinearity.
In \sect{sec:phonon}, we study hBN with phononic nonlinearity under infrared pump and show the Floquet phonon-polaritons in both hBN monolayers and thin flakes.
In \sect{sec:jplasmon}, we turn to layered superconductors driven by THz fields along the out-of-plane direction and predict Floquet Josephson plasmons that appear as pump-induced resonances in the probe reflectivity.
Technical details, including the Keldysh formulation, generalized Green functions for slab geometries, and derivations for the graphene and hBN examples, are provided in the Appendices.

\section{General formalism}
\label{sec:formalism}

\begin{figure}
	\includegraphics[width=\linewidth]{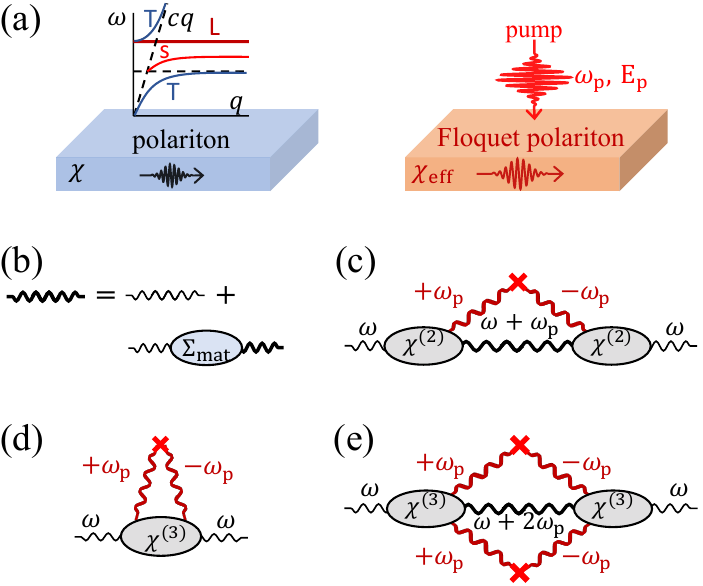}
	\caption{\label{fig:theoryFig} 
		(a) 
		Left: polaritons in a solid state material in equilibrium with the optical susceptibility $\chi$. 
		The plot is a schematic  dispersion of the bulk and surface polaritons in a three dimensional isotropic polar insulator, with ``T'', ``L'' and ``s'' labeling transverse bulk modes, longitudinal bulk modes and surface modes, respectively.
		Right: a strong pump field (red wavy line) drives the system and modifies the susceptibility to $\chi_{\text{eff}}$ via optical nonlinearity, transforming the polaritons to Floquet polaritons.
		(b) The thick/thin black wavy lines denote the bare polariton propagator $G_0$/photon propagator.
		The material corrections are incorporated through a Dyson expansion by contributing a self-energy $\Sigma_{\text{mat}}=\chi$, which is just the equilibrium susceptibility.
		(c)  Self energy (effective susceptibility) contributed by the pump via  $\chi^{(2)}$.
		The red wavy line denotes the pump field.
		(d, e) Pump-contributed self-energies arising from $\chi^{(3)}$.
		Note that for single-frequency coherent pump (i.e., from a laser), 
		the connection of $\omega_{\text{p}}$ and  $-\omega_{\text{p}}$ pairs of the pump lines by a cross sign is arbitrary. 
		It is shown as a guide to the eye for selecting the polaritonic self-energies, i.e., the 1PI diagrams which cannot be separated into halves by cutting a polariton propagator at frequency $\omega$.
		Therefore, only non-crossing diagrams are involved.
		For incoherent pump, crossing diagrams also contribute~\cite{xiao2025interactioneffectselectronicfloquet}, similar to the case of disorder scattering.
	}
\end{figure}

\subsection{Polaritonic propagators}
The Floquet polaritons in optically driven materials can be  captured by the Maxwell's equations for electromagnetic (EM) waves combined with the linear and nonlinear optical susceptibilities of the material.
The full quantum formalism is most conveniently expressed in terms of the Keldysh Green functions~\cite{keldysh1964diagram,schwinger1961brownian,altland2010condensed,kamenev2023field}, which we present in Appendix~\ref{app:keldysh}. 
Since its classical saddle points are extremely good approximations in most experimental situations, we show its classical version in the main text. 
Specifically, the Maxwell's equations lead to the dynamics of  the electric field as
\begin{align}
\nabla\times(\nabla\times\mathbf{E})=&  -\frac{1}{c^2}\partial_t^2(\mathbf{E}+4\pi\mathbf{P}), 
\label{MW}
\end{align}
where the electric polarization $P$ contains the material degrees of freedom.
It is related to the \textit{net} electric field as 
\begin{align}\label{pe}
 P_{\text{i}}(\omega)= 	& \chi_{\text{ij}}(\omega) E_{\text{j}}(\omega)
+ \sum \chi^{(2)}_{\text{ijk}}(\omega_1, \omega_2) 
	E_{\text{j}}(\omega_1)E_{\text{k}}(\omega_2)
\notag\\
	&
	+ \sum \chi^{(3)}_{\text{ijkl}}(\omega_1, \omega_2, \omega_3) 
	E_{\text{j}}(\omega_1)
E_{\text{k}}(\omega_2) E_{\text{l}}(\omega_3)
\notag\\
	& +\dots \,.
\end{align}
where we have used the compact symbol $\omega$ to represent the frequency-momentum $(\omega, q)$ for notational simplicity. 
Here $\chi$ is the linear optical susceptibility tensor, $\chi^{(n)}$ is the nonlinear susceptibility tensor of the $n$th order, summation over repeated indices is assumed, and the explicit summations run over the frequencies and momenta with the constraint $\sum_a \omega_a=\omega,\, \sum_a q_a=q$. 
We have focused on the most common case of polaritons arising from the coupling between light and electric dipoles. Examples include plasmons, phonon polaritons, exciton polaritons, etc.
While the polaritons from magnetic dipoles, e.g., magnon polaritons, are not explicitly included in \equa{MW} and \equa{pe}, their description is straightforward by analogy.

At the level of linear response, the polariton spectrum is fully determined by the linear susceptibility $\chi(\omega,\mathbf{k})$. By substituting the linear polarization $\mathbf{P}(\omega,\mathbf{k})=\chi(\omega,\mathbf{k})\mathbf{E}(\omega,\mathbf{k})$ into Maxwell's equation Eq.~(\ref{MW}), one obtains  the following relation:
\begin{equation}
[\omega^2(1+4\pi\chi(\omega,\mathbf{k}))-c^2k^2]\mathbf{E}(\omega,\mathbf{k})=4\pi \partial_t^2 \mathbf{P}_{\mathrm{ext}}.
\label{eq:defPolariton}
\end{equation}
where we have kept a possible external  polarization source $\mathbf{P}_{\mathrm{ext}}$, and have neglected the off-diagonal components of $\chi(\omega,\mathbf{k})$ for simplicity.
Its general solutions define the electromagnetic eigenmodes in the material.
We define the retarded Green function (propagator) 
\begin{align}\label{eqn:G_0}
G_0(\omega,\mathbf{k})=\frac{-4\pi\omega^2}{\omega^2(1+4\pi\chi(\omega,\mathbf{k}))-c^2k^2}
,
\end{align}
for electric fields,  as shown in \fig{fig:theoryFig}(b). The prefactor $-4\pi\omega^2$ is chosen such that the electric field generated by an external source can be written as $\mathbf{E}_{\mathrm{r}}=G_0\mathbf{P}_{\mathrm{ext}}$.
The poles of $G_0(\omega,\mathbf{k})$ correspond to polariton branches.
Through the $\chi$ term, the material response takes effect via the self-energy correction  $\Sigma_{\text{mat}}= \chi$ to the photon propagators.


More generally, the formalism is not restricted to transverse bulk modes. 
For longitudinal bulk modes, surface polaritons, or hyperbolic bulk polaritons, one should generalize the propagator $G_0$ in \equa{eqn:G_0} by the appropriate component of the linear electromagnetic Green function, determined by Maxwell equations taking into account the anisotropy of the susceptibility and the geometry and boundary conditions of the system. 
For example, for a two-dimensional (2D) system embedded in vacuum, one should use
\begin{equation}\label{eq:G_0nearfield}
G_0^{\mathrm{nf}}(\omega,\mathbf{q})
=
\frac{-2\pi q}{1+2\pi q \chi(\omega,\mathbf{q})},
\end{equation}
in the quasi-static (near-field) limit, where $\chi$ is the two dimensional version of the optical susceptibility. 
Note that the $\chi$ is defined as the response polarization divided by the perturbing electric field, which should be distinguished from the one frequently defined as the response charge density divided by the perturbing electrical potential.
Again, the definition of the propagator is chosen so that the response electric field on the 2D plane is related to an external polarization source on the plane as $\mathbf{E}_{\mathrm{r}}=G_0^{\mathrm{nf}} \mathbf{P}_{\mathrm{ext}}$.
The poles of $G_0^{\mathrm{nf}}$ correspond to longitudinal surface-plasmon or phonon-polariton modes. 
For a derivation, see Appendix~\ref{app:slabgreen} which also contains the 
formalism for polaritons in anisotropic media and slab geometries that includes hyperbolic polaritons.
With this replacement, all subsequent constructions of $\chi_{\mathrm{eff}}$ and the dressed Green function remain unchanged.

\subsection{Pump induced effective susceptibility}

The pump contributes additional self-energies $\Sigma$ to the polariton propagator $G$ in the form of a Dyson equation:
\begin{equation}
	G^{-1}(\omega,\vec{k})=G^{-1}_0(\omega,\vec{k})-\Sigma(\omega,\vec{k}),
	\label{eq:dyson}
\end{equation}
where the self-energy correction $\Sigma$ arises from the nonlinear  optical effects combined with the pump. 
The poles of the dressed propagator $G$ gives the dispersion	 of the Floquet polaritons.
Viewing the polaritons and the pump light as interacting bosons, this self-energy $\Sigma(\omega,\vec{k})$ admits a diagrammatic expansion  where each term corresponds to a physical process involving one or more nonlinear susceptibilities combined with the pump, and/or internal electric field propagators, see Fig.~\ref{fig:theoryFig}(c)(d)(e). 
For example, 
Fig.~\ref{fig:theoryFig}(d) depicts the pump-contributed self-energy involving a single vertex associated with the third-order nonlinear susceptibility $\chi^{(3)}$,
while Fig.~\ref{fig:theoryFig}(e) contains two $\chi^{(3)}$  vertices involving internal polariton propagators. 
Fig.~\ref{fig:theoryFig}(c) shows a similar contribution from the second order nonlinear susceptibility $\chi^{(2)}$. 
These self-energies may be viewed as pump-contributed  linear susceptibilities that add to the total effective susceptibility as
\begin{align}
\chi_{\text{eff}}(\omega,\vec{k})=  \chi(\omega,\vec{k}) + \Sigma(\omega,\vec{k}).
\end{align}
They are conveniently expanded in powers of the pump field and nonlinear optical susceptibilities following the guidance of the diagrams in \fig{fig:theoryFig}.

The full quantum formalism for this perturbative expansion in the language of Keldysh path integral~\cite{keldysh1964diagram,schwinger1961brownian,altland2010condensed,kamenev2023field} is shown in Appendix~\ref{app:keldysh}.
That approach also takes into account the corrections from quantum/thermal fluctuations, whose effects are the goal of equilibrium perturbation theory in textbooks.
For example, if one replaces the two pump lines in   \fig{fig:theoryFig}(d) by a polariton propagator, this self-energy means that the  quantum/thermal fluctuations of the polaritons would shift the single polariton energy via the third-order nonlinearity, or in other words, via two-polariton interaction.
These fluctuation effects are generally weak for polaritons except for certain species such as exciton polaritons~\cite{Carusotto2013QuantumFluids,
	Savvidis2000AngleResonantStimulatedPolaritonAmplifier,
	Saba2001HighTemperatureParametricAmplification,
	Barachati2018InteractingPolaritonFluids,
	Wu2021NonlinearParametricScattering,
	Zhao2022NonlinearPolaritonParametricEmission,
	Xiang2026DipolarPolaritons}.
In the presence of the strong pump, the dominating diagrams are therefore those  with all fluctuation lines replaced by the classical pump, as in \fig{fig:theoryFig}(c)(d)(e). 
In the main text, we derive the Floquet polaritons  at the classical level, which can be done  simply from solving \equa{MW} and  \equa{pe} iteratively.

Consider a system with a nonzero ($l+1$)-order susceptibility $\chi^{(l+1)}$ and in the presence of a strong pump field $2\mathbf{E}_{\text{p}}\cos(\omega_{\text{p}}t)$, the nonlinear polarization $\mathbf{P}_{\text{NL}}(\omega',\mathbf{k}')$ from higher powers of the pump and probe fields in \equa{pe} appears on the right-hand side of  \equa{MW}.
This indicates that Eq.~(\ref{eq:defPolariton}) alone is no longer sufficient to determine the polariton branches under a strong pump. 
Physically, polaritons are self-sustained oscillations of the weak probe  electric field $\mathbf{E}(\omega,\mathbf{k})$ that can propagate in the system without requiring an external source, as shown in Fig.~\ref{fig:theoryFig}(a). 
Its linear  propagator  reflects the intrinsic properties of collective excitations in the driven system, i.e., the Floquet polaritons. 

To obtain the Floquet polariton propagators, we collect terms in $\mathbf{P}_{\text{NL}}$ that are linear in $E_{\text{pr}}$:
\begin{align}
\mathbf{P}_{\mathrm{NL}} &= \sum_l \chi^{(l+1)}( \mathcal{P}\{\pm\omega_{\text{p}}, \ldots, \pm\omega_{\text{p}},\omega_{\text{pr}}\}) \left( \mathbf{E}_{\text{p}} \right)^l \mathbf{E}_{\text{pr}},
\notag\\
&  = \sum_{l, n} 
e^{-i n\omega_{\text{p}} t}
\tilde{\chi}^{(l+1)}_{{n}}(\omega_{\text{pr}}) 
\left( \mathbf{E}_{\text{p}} \right)^l
\mathbf{E}_{\text{pr}}.
\label{eq:interpe}
\end{align}
Here $\mathcal{P}$ denotes all distinct permutations of the input frequency arguments whose sum gives the frequency of the nonlinear polarization arising from multi-photon mixing between the pump and probe. 
Unlike \equa{pe}, the spatial indices have been suppressed here for notational simplicity.
In the second line of Eq.~(\ref{eq:interpe}), we have separated out the components of  $\mathbf{P}_{\mathrm{NL}}$ at the frequencies $\omega_{\text{pr}}+n\omega_{\text{p}}$, and have 
defined  $\tilde{\chi}^{(l+1)}_{{n}}(\omega)$ where $n$ labels the Fourier (Floquet) index. 
Each component acts as a driving term 
$ e^{-i n\omega_{\text{p}} t} \left( \mathbf{E}_{\text{p}} \right)^l \mathbf{E}_{\text{pr}}$ 
 in \equa{MW} that radiates fields $\mathbf{E}_{\text{r}}$ at the frequency $\omega_{\text{pr}}+n\omega_{\text{p}}$.
 \fig{fig:theoryFig}(d) is an example of $n=0$, a static `driving' term.
In \fig{fig:theoryFig}(c) and  \fig{fig:theoryFig}(e), if one cuts the diagrams vertically into halves, the left halves represent an $n=1$ and $n=2$ term, respectively. 
 Plugging the polarization $\mathbf{P}(\omega,\mathbf{k})=\chi(\omega,\mathbf{k})\mathbf{E}(\omega,\mathbf{k})+\mathbf{P}_{\mathrm{NL}}$ into \equa{MW} gives the linear differential equation $\hat{L} \mathbf{E}_{\text{pr}}=0$ for the periodically driven polaritons.
 In principle, one may obtain the Floquet polariton spectrum by diagonalizing $\hat{L}$ in frequency space numerically with a cutoff in the Floquet replica index.
 However, we proceed with a more physically transparent approach in the following.
 
Because the continuous translation symmetry in time is reduced to a discrete one in the pumped material,  the  Floquet polariton propagators $G(\omega+m\omega_{\text{p}},\omega)$ are in principle functions of two frequencies where $m$ is an integer.
It is the inverse of the linear Floquet matrix $\hat{L}$.
Physically, this means that in a pump-probe experiment, the response to the probe field also contains frequency components $\mathbf{E}_{\text{pr}}+ m \omega_{\text{p}}$.
The propagator in \equa{eq:dyson} is the component $G(\omega) \equiv G(\omega,\omega)$  that preserves the input frequency, which is our focus in this paper and is what most experiments measure.
For the propagator $G(\omega)$, the polaritonic self-energies  $\Sigma(\omega)$ are therefore the one-particle-irreducible (1PI) diagrams which cannot be separated  by cutting a polariton propagator at frequency $\omega$, see \fig{fig:theoryFig}(c)(d)(e).
This self-energy, or in other words, the  pump-contributed effective linear susceptibility, can be built from $\tilde{\chi}^{(i+1)}_{{n}}$  as
\begin{align}\label{eq:chieff}
\Sigma(\omega) =
& \sum_{i} \tilde{\chi}^{(i+1)}_{n=0}(\omega) 
	\, \mathbf{E}_{\text{p}}^{i}  +
		 \sum_{ij,n\neq 0}
	\bigg[ \tilde{\chi}^{(j+1)}_{-n}(\omega + n\omega_{\text{p}}) \,
\notag
\\
& 	\mathbf{E}_{\text{p}}^{j} G_0(\omega + n\omega_{\text{p}}) \,
	\tilde{\chi}^{(i+1)}_{n}(\omega) \mathbf{E}_{\text{p}}^{i} \bigg]
+\cdots.
\end{align}
The $n=0$ components of $\mathbf{P}_{\mathrm{NL}}(\omega_{\text{pr}})$ contribute time-translational-invariant terms 
$\tilde{\chi}^{(i+1)}_{{0}} \left( \mathbf{E}_{\text{p}} \right)^i \mathbf{E}_{\text{pr}}$ in \equa{MW}, which are therefore already effective linear susceptibilities from the perspective of the probe field. 
\fig{fig:theoryFig}(d) is a typical example. 
On the contrary, the $n \neq 0$ components radiate fields at shifted frequencies: $\mathbf{E}_{\text{r}}(\omega_{\text{pr}}+n\omega_{\text{p}})=G_0(\omega_{\text{pr}}+n\omega_{\text{p}}) \mathbf{P}_{\mathrm{NL}}(\omega_{\text{pr}}+n\omega_{\text{p}})$
where $G_0$ is from \equa{eqn:G_0}. 
This radiated field can then generate a nonlinear polarization component $\sum_j \tilde{\chi}^{(j+1)}_{\text{m}}(\omega_{\text{r}}) \left( \mathbf{E}_{\text{p}} \right)^j \mathbf{E}_{\text{r}}$ through a further nonlinear process. 
For $m=-n$, this process brings the response back to the original probe frequency $\omega_{\text{pr}}$, 
thereby contributing an effective  linear susceptibility as  the second term in \equa{eq:chieff}. 
For example, \fig{fig:theoryFig}(c) shows the effective susceptibility from an $n=1, i=1$ term in \equa{eq:chieff}, 
while \fig{fig:theoryFig}(e) shows an $n=1, i=2$ term. 
The generalization of \equa{eq:chieff} to a slab geometry and to a 2D sheet is shown in \append{app:slabgreen}.

The dressed Green function $G(\omega,\mathbf{k})$ is obtained by plugging the self-energy in \equa{eq:dyson}, or equivalently,  by replacing $\chi$ with $\chi_{\text{eff}}$   in \equa{eqn:G_0}. 	
Floquet polariton branches are then identified as the poles of the dressed Green function.
One immediately observes that  whenever $\omega+n\omega_{\text{p}}$ matches the frequency of an intrinsic polariton, 
a resonant enhancement in $\chi_{\text{eff}}$   occurs from the pole of $G_0(\omega+n\omega_{\text{p}})$. 
This implies the emergence of new Floquet polariton branches at closeby frequencies.

The remaining terms in \equa{eq:chieff} shown as `$\cdots$'  are higher order effects in the Floquet driving terms, which generate nested diagrams.
For example, the polariton propagator in \fig{fig:theoryFig}(c)(e) can have corrections from the pump too.
Note that for single-frequency coherent pump, crossing diagrams are absent because at all orders of the pump, one can always pair positive and negative frequency pump lines in a way that no crossing occurs, see SI in Ref.~\cite{xiao2025interactioneffectselectronicfloquet}.
However, this does not mean that the exact self-energy can be obtained from replacing the polariton propagator in \fig{fig:theoryFig}(c)(e) by the dressed Green function $G(\omega,\mathbf{k})$, as 
this series includes some diagrams that are not 1PI by cutting a polariton propagator at frequency $\omega$.

In the following, we apply this method to predict the Floquet polariton spectra in several concrete material platforms that have been under intensive experimental investigation.

\section{\label{sec:gplasmon}Floquet Graphene plasmons}
\begin{figure}
	\includegraphics[width=\linewidth]{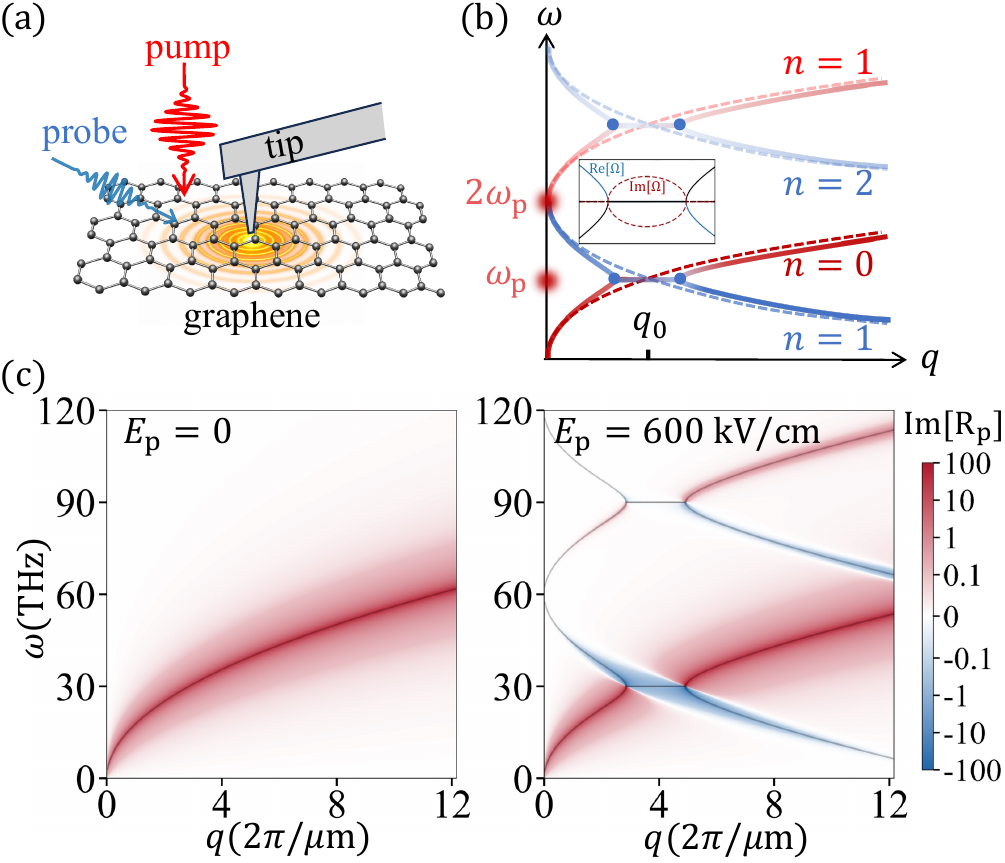}
	\caption{\label{fig:GrPlasmon} 
		Floquet plasmons in graphene.
		(a) Schematic of the near-field pump-probe setup for creating and measuring Floquet plasmons in graphene.
		(b) Schematic dispersion of Floquet 2D plasmons under strong optical pumping with the photon frequency $\omega_{\text{p}}$. 
		The original branch (black dashed curve labeled by $n=0$) and its Floquet replicas (dashed curves labeled by  indices $n=0,1,2$) shifted by $2n\omega_{\text{p}}$ intersect and hybridize into the Floquet plasmons (solid curves).
		The blue dots are exceptional points. 
		The inset is the schematic behavior of the real and imaginary parts of the plasmon frequency relative to $\omega_{\text{p}}$ as a function of momentum near the band-crossing point.
		(c) Imaginary parts of the near-field reflection coefficient $R_{\text{p}}(\omega, q)$ of graphene  pumped with the field $E_{\text{p}}=0$ (left panel) and $E_{\text{p}}=600 \unit{kV/cm}$ (right panel, showing the Floquet plasmons).
		It is plotted on the frequency-momentum plane with $q$ parallel to the linearly polarized pump field.
		The color scale is on a logarithmic scale so that the plasmons appear broader than what they actually are.
		The solid curves are real parts of the plasmon frequencies.
		The parameters used are the pump photon frequency $\omega_{\text{p}}=30 \unit{THz}$, the carrier density $n=6.5 \times 10^{12}\unit{cm}^{-2}$ corresponding to $\varepsilon_\text{F}= 300 \unit{meV}$,
		the Fermi velocity $v_{\text{F}}=10^8 \unit{cm/s}$,  and the intrinsic plasmonic damping rate $\gamma=1.5 \unit{THz}$ 
		motivated by experiments~\cite{Ni2016UltrafastGraphenePlasmons, Ni2018FundamentalLimitsGraphenePlasmonics}. 
	}
\end{figure}

The first example is the two-dimensional plasmons in graphene pumped by far field laser in the mid and far infrared regime. 
The Dirac dispersion of electrons  breaks Galilean invariance, giving rise to a strong
third-order nonlinear susceptibility $\chi^{(3)}$~\cite{mikhailov2007non,
	Hendry2010GrapheneNonlinear,
	Kumar2013GrapheneTHG, 
	cheng2014third,
	Rostami2017PlasmonicEffectsNonlinearOpticsGraphene,
	sun2018third,
	Jiang2018GateTunableGrapheneNonlinear}.
This nonlinearity combined with the pump naturally transforms the collective electronic excitations to  Floquet plasmons, see \fig{fig:GrPlasmon}(b) for a schematic dispersion.

At the frequencies $\omega \ll 2\varepsilon_{\text{F}}$ and in-plane momenta $q \ll \omega/v_\text{F}$ where $\varepsilon_\text{F}$ ($v_\text{F}$) is the Fermi energy (Fermi velocity), the  two dimensional optical  conductivity is well approximated by the Drude form $\sigma=iD/[\pi (\omega+i\gamma)]$ where $D=v_{\text{F}}k_{\text{F}}e^2/\hbar$ is the Drude weight, $k_{\text{F}}$ is the Fermi momentum, and $\gamma$ is the momentum-relaxing scattering rate.
The corresponding  two dimensional optical susceptibility  is $\chi=-D/[\pi\omega (\omega+i\gamma)]$.
Plugging it into \equa{eq:G_0nearfield}, one obtains the  bare Green function for the plasmons:
\begin{equation}\label{eq:G_0_plasmon}
	G_0^{\mathrm{pl}}(\omega,\mathbf{q})
	=
	\frac{-2\pi q}{1- 2 q D/[\omega (\omega+i\gamma)]}.
\end{equation}
Therefore, the intrinsic plasmon frequency at momentum $q$ is $\omega_{\text{q}}-i\gamma/2$ which has the real part $\omega_{\text{q}} \approx \sqrt{2Dq}$, see black dashed curve in Fig.~\ref{fig:GrPlasmon}(b). 
Note that there are also branches at negative frequencies and momenta $(\pm\omega_{\text{q}},\pm q)$, not just  that shown in Fig.~\ref{fig:GrPlasmon}(b). 

\subsection{Floquet plasmons from third-order nonlinearity}
The third-order susceptibility contributed by electronic intraband effects is~\cite{sun2018third} 
\begin{align}\label{eq:chi3}
\chi^{(3)}_{ilmn}(\omega_1, \omega_2,\omega_3)=-\frac{ D^{(3)} }{ (\omega_1+ \omega_2 + \omega_3) \omega_1 \omega_2 \omega_3} \Delta_{ilmn}
\end{align}
where $D^{(3)}=v_{\text{F}}e^4/(24\pi\hbar^3k_{\text{F}})$ is the ``third-order Drude weight'' at zero temperature in the kinetic regime and  $\Delta_{ilmn}=\delta_{il}\delta_{mn}+\delta_{im}\delta_{ln}+\delta_{in}\delta_{lm}$ with all indices referring to spatial components.
When the system is driven by a  normally incident linearly polarized pump $\mathbf{E}_{\text{p}}=E_{\text{p}}\sin(\omega_{\text{p}} t)\hat{x}$ corresponding to the vector potential $\mathbf{A}_{\text{p}}=A_{\text{p}}\cos(\omega_{\text{p}} t)\hat{x}$,
 it combined with $\chi^{(3)}$ contributes via \fig{fig:theoryFig}(d)(e) to 
 the effective susceptibility $\chi_{\text{eff}}(\omega, q)$.
The latter is no longer isotropic as it depends on the direction of $q$ and probing field.
In the following, we focus on the probing momenta along $\hat{x}$, which means the probe field is also along $\hat{x}$ since the plasmons are longitudinal modes.

Taking care of the spatial indices and the possible combination of frequency arguments in \equa{pe}, the self-energy in  \fig{fig:theoryFig}(d) reads
\begin{align}
	\chi^{(d)}(\omega,\mathbf{q})
	=\frac{9  D^{(3)} }{2 \omega^2 \omega_{\text{p}}^2}  E_{\text{p}}^2
    = \frac{3}{16} \xi^2 \frac{ D }{\pi \omega^2}
	,
	\label{eq:sigmae_plasmon}
\end{align}
where $\xi=e E_{\text{p}}/(\hbar k_{\text{F}} \omega_{\text{p}})$ is a dimensionless measure of the pump strength, and we have made use of the expressions 
$D=v_{\text{F}}k_{\text{F}}e^2/\hbar$ 
and $D^{(3)}=v_{\text{F}}e^4/(24\pi\hbar^3k_{\text{F}})$ of the first  and third-order optical weights.
Apparently, this self-energy  simply shifts the effective linear Drude weight from $D$ to a lower value: $\tilde{D}=D(1-\frac{3}{16} \xi^2)$~\cite{sun2018third}.

Similarly, the self-energy in  \fig{fig:theoryFig}(e) reads
\begin{align}
	\chi^{(e)}(\omega,\mathbf{q}) = &
	 \chi^{(3)}  
	G_0^{\mathrm{pl}}(\omega+2\omega_{\text{p}},\mathbf{q}) 
	\chi^{(3)} E_{\text{p}}^4 
	+ (\omega_{\text{p}} \rightarrow -\omega_{\text{p}})
\notag \\
	= &
	\frac{2\pi q \left(\frac{3 \xi^2 D}{16\pi \omega}\right)^2}{2q\tilde{D}-(\omega+2\omega_{\text{p}})^2}
+ (\omega_{\text{p}} \rightarrow -\omega_{\text{p}})
\,.
\label{eq:sigma_plasmon}
\end{align}
In the first term, proper frequency arguments and contraction of spatial indices (see \equa{eq:chi3}) are presumed, and we have replaced the Drude weight in the plasmon propagator by $\tilde{D}$, corresponding to adding the self-energy \fig{fig:theoryFig}(d) to  the  plasmon propagator in  \fig{fig:theoryFig}(e).
Physically, the probe field (plasmonic field) $E_{\text{pr}}(\omega,q)$ with two pump fields $E_{\text{p}}(\omega_{\text{p}})$ induces an electrical polarization $P(\omega+2\omega_{\text{p}},q)$ through $\chi^{(3)}$.
Afterwards, the polarization emits an electric field $E(\omega+2\omega_{\text{p}},q)$, which with two negative-frequency  pump fields $E_{\text{p}}(-\omega_{\text{p}})$ induces the polarization $P(\omega,q)$.
The latter is at the same frequency and momentum as the probe field and is linear in it.
Therefore, this process contributes an effective linear susceptibility seen by the plasmonic field.

Collecting these two self-energies and the intrinsic linear susceptibility $\chi$, the total effective susceptibility is found as
\begin{align}
	\chi_{\text{eff}}(\omega,\mathbf{q})=
	&-\frac{\tilde{D}}{\pi\omega(\omega+i\gamma)}
	 + 
	\frac{\frac{D}{\pi} \frac{2 q D}{\omega^2}	\left(\frac{3  \xi^2}{16}\right)^2}
	{2q\tilde{D}-(\omega \pm 2\omega_{\text{p}})^2}
.
\label{eq:grchieff}
\end{align}
Here the `$\pm$' in the second term means there are actually two terms, one with `$+$' and one with `$-$', corresponding to two diagrams with different signs of pump frequencies, see \equa{eq:sigma_plasmon}.
For notational simplicity, we will occasionally use this convention in the rest of the paper such as in Eqs.~\eqref{eq:chieffhbn} and \eqref{eq:JPeeff}.
Note that with a nonzero scattering rate, the $(\omega \pm 2\omega_{\text{p}})^2$  in \equa{eq:sigma_plasmon} should be replaced by $(\omega \pm 2\omega_{\text{p}})(\omega \pm 2\omega_{\text{p}}+i\gamma)$.
The Floquet polariton branches are found as the poles of  the propagator 
\equa{eq:G_0nearfield} with the $\chi$ replaced by the $\chi_{\text{eff}}$ in \equa{eq:grchieff}.
Equivalently,  they are determined by the condition of vanishing 2D dielectric: $\epsilon_{\text{2D}} = 1 +2\pi q\chi_{\text{eff}}=0$. 
From the structure of \equa{eq:grchieff}, one observes that Floquet replicas of plasmon branches appear, shifted vertically by integer multiples of $2\omega_{\text{p}}$, see Fig.~\ref{fig:GrPlasmon}(b). 
Because only the lowest order diagram in \fig{fig:theoryFig}(e) is calculated, only  the $n=\pm 1$ replicas are included in \equa{eq:grchieff}. 
Nesting the self-energy in \fig{fig:theoryFig}(e) to the propagator in itself would result in more replicas, which we neglect for simplicity and without loss of the essential physics.

Note that in equilibrium, $\mathrm{Im}[\chi]=\mathrm{Re}[\sigma]/\omega$  is guaranteed to be positive at positive frequencies (and negative at negative frequencies), which is a consequence of causality and stability.
However, the pole at $-\omega_q+2\omega_{\text{p}}$ in \equa{eq:grchieff}  has negative spectral weight ($\mathrm{Im}[\chi_{\text{eff}}(-\omega_q+2\omega_{\text{p}}, q)]<0$)
because it comes from  shifting the negative frequency plasmon  at $-\omega_q$ up by $2\omega_{\text{p}}$ to positive frequency.
This means the system does not absorb energy from external perturbation in this frequency-momentum range but exports energy to it, a hallmark of non-equilibrium phenomena.
We will go back to this point in \sect{sec:flat_band}.
Furthermore, because the coefficient $|\chi^{(3)}|^2 E_{\text{p}}^4$ of the shifted pole is always positive, its negative spectral weight  is a generic result, which occurs also for the other examples such as Floquet phonon polaritons in \sect{sec:phonon} and Floquet Josephson plasmons in \sect{sec:jplasmon}.

\subsection{Plasmonic flat band and momentum gap}
\label{sec:flat_band}
When a Floquet-shifted branch intersects with the original one, for example at $(\omega_{\text{q}_0}, q_0)$ with $2\tilde{D} q_0 = \omega_{\text{p}}^2$, strong hybridization occurs between the $n=0$ and $n=1$ branches, as shown in Fig.~\ref{fig:GrPlasmon}(b).
This effect manifests as the resonant pole at $\omega= \omega_{\text{p}}$ of the third term in \equa{eq:grchieff}.
It occurs because the plasmon at $(\omega_{\text{q}_0}= \omega_{\text{p}}, q_0)$  is dressed by two pumps with frequency $-\omega_{\text{p}}$ through the third-order nonlinearity to yield a polarization at frequency $-\omega_{\text{p}}$, which resonantly couples to the plasmon mode at $(-\omega_{\text{q}_0}, q_0)$, as shown by \fig{fig:theoryFig}(e).
Near $(\omega_{\text{q}_0}, q_0)$, the polariton spectrum is expected to exhibit nontrivial behavior.
In the following, we substitute Eq.~(\ref{eq:grchieff}) into the polariton condition $\epsilon_{\text{2D}} = 1 +2\pi q\chi_{\text{eff}}=0$ to study this behavior.

Close to $(\omega_{\text{p}}, q_0)$, defining the dimensionless small de-tunings of the frequency and momentum as $\omega=\omega_{\text{p}}(1+\Omega)$ and $2\tilde{D} q=\omega_{\text{p}}^2(1+Q)^2$, one may keep the leading order behavior of  \equa{eq:grchieff} and approximate $\epsilon_{\text{2D}}$ as 
\begin{align}
\epsilon_{\text{2D}}
\approx 
2 (\Omega-Q+i\gamma'/2)+ \frac{2\kappa^2}{\Omega+Q+i\gamma'/2}
\label{eq:epsilon}
\end{align}
where $\kappa = 3 \xi^2/32$ is a dimensionless `coupling strength' and $\gamma' = \gamma/\omega_{\text{p}}$.
Note that we have kept only the first and third terms in $\chi_{\text{eff}}$ since these two terms dominate close to $(\omega_{\text{q}_0}, q_0)$ and already contain the essential physics of coupling between the $n=0$ and $n=1$ branches in  Fig.~\ref{fig:GrPlasmon}(b). 
From the polariton condition $\epsilon_{\text{2D}}=0$,  one obtains the eigenmode equation 
\begin{equation}
\left(\Omega + i \frac{\gamma'}{2} \right)^2
=Q^2 -\kappa^2
\,.
\label{eq:momentumgaprange}
\end{equation}
The eigenfrequencies are thus found as 
\begin{equation}
\omega_q=  \omega_{\text{p}} 
\left(1- i \frac{\gamma'}{2} \pm \sqrt{Q^2-\kappa^2}
\right)
\,.
\label{eq:floquetplasmon_freq}
\end{equation} 
	
\equa{eq:floquetplasmon_freq} shows typical behaviors of non-Hermitian coupling between two modes.
At large momentum  detuning ($|Q| > \kappa$) from $q_0$, the two hybridized modes have different real parts of frequency.
At small detuning ($|Q| < \kappa$), they have the same real parts but are split in imaginary parts.
The two points at $|Q| = \kappa$ are exceptional points where the two Floquet modes are exactly degenerate in the complex frequency, see Fig.~\ref{fig:GrPlasmon}(b). 
Therefore, a nearly perfectly flat plasmonic band is created in the momentum range between the two exceptional points.
This feature in band dispersion is also called a `momentum gap'.
The non-Hermitian coupling comes from the fact that the original plasmonic pole at $\sqrt{2\tilde{D} q}$ has positive spectral weight, while the replica pole at
$-\omega_q+2\omega_{\text{p}}$ in \equa{eq:grchieff} has negative spectral weight, making their hybridization `attractive', 
contrary to the  normal case of two same-sign poles having `repulsive' hybridization between modes.
Note that the apparent `non-Hermitian' effects here is created from the periodic drive instead of a bath, showcasing that non-Hermitian coupling  could be created simply by pumping a closed system.

In the small detuning region ($|Q| < \kappa$) and in the ideal case of zero intrinsic damping, one of the two Floquet plasmons has negative damping.
Physically, this corresponds to a parametric down-conversion process, where two pump photons convert into two plasmon-polaritons with equal frequency but opposite momentum, leading to parametric instability of this mode~\cite{Sun_plasmon.2022,kiselev2024inducing}, see \sect{appendix:parametric instability} for an equation-of-motion description. 
Because of the nonzero intrinsic damping $\gamma$, parametric instability  for  the mode at zero detuning occurs only if the pump exceeds a threshold set by $\kappa=\gamma'/2$, meaning $\xi>4\sqrt{{\gamma}/({3\omega_{\text{p}}})}$~\cite{Sun_plasmon.2022}.
For a numerical estimation, taking the  damping rate $\gamma/\omega_{\mathrm p}=0.05$ used in Fig.~\ref{fig:GrPlasmon}, 
the threshold is $\xi_{\mathrm{c}}\simeq 0.52$. 
At the carrier density $n=6.5 \times 10^{12}\unit{cm}^{-2}$ and pump frequency $\omega_{\mathrm p}=30\unit{THz}$, the corresponding threshold pump field is found as
$
E_{\mathrm{c}}=
\xi_{\mathrm{c}} \hbar k_{\text{F}} \omega_{\text{p}}/e
\approx  290 \unit{kV/cm}
$.
If pumped with a field beyond the threshold, the Floquet plasmon modes around $q_0$ grow exponentially after being excited.
Therefore, the system exhibits plasmonic gain and may function as a plasmonic repeater and amplifier.
In the quantum mechanical sense, these modes become squeezed states whose quantum feature can be measured by homodyne experiments~\cite{RevModPhys.81.1727,Sun_plasmon.2022,dapolito2026quantumlightnanoimaging},
making these systems potential quantum plasmonic repeaters and sources of entangled plasmons~\cite{Sun_plasmon.2022}.
Of course, in continuous-wave experiments, this unstable growth is expected to be cut off by nonlinear saturation that is beyond the present weak-probe treatment.

In near field experiments, the non-equilibrium effects and parametric instability manifest as negative imaginary parts of the near field reflection coefficient:
\begin{align}\label{eqn:Rp}
R_{\text{p}}= 1- \frac{1}{\epsilon_{\text{2D}}}
\end{align}
defined as the ratio between the reflected electric potential and the incident one by the 2D sample~\cite{sun2020collective}.
It is named `$R_{\text{p}}$' because it is the near field limit ($q \gg \omega/c$) of the Fresnel reflection coefficient of the p-polarized light.
Note that in equilibrium, causality and stability constrain $\mathrm{Im}[R_{\text{p}}]$  to be positive, meaning that the sample absorbs energy from the external perturbation.
The right panel of \fig{fig:GrPlasmon}(c) shows $\mathrm{Im}[R_{\text{p}}]$  of graphene pumped at a certain field strength and frequency.
Near the band crossing region, $\mathrm{Im}[R_{\text{p}}]$ shows typical non-equilibrium features  with both positive and negative signs, whose  structure can be understood analytically by plugging \equa{eq:epsilon} into \equa{eqn:Rp}.
The region with negative sign indicates that the sample works as a gain medium, exporting energy to the external perturbation.
Specifically, the flat band region has negative sign because the pole of the unstable mode (the mode with `$+$' sign in \equa{eq:floquetplasmon_freq}) is closer to the real axis, and thus dominates the response.
Moreover, the  $\mathrm{Im}[R_{\text{p}}]$ is all negative over the lower half of the $n=1$ replica, which is a consequence of the negative spectral weight of the $-\omega_q+2\omega_{\text{p}}$ pole in  \equa{eq:grchieff}.

\subsection{Parametric oscillator picture}
\label{appendix:parametric instability}
The Floquet spectra and parametric instability of the plasmon modes at $\omega_{\text{p}}$ can  be 
directly understood from their equation of motion under pump.
Within the quasi-static (near-field) approximation, the Lagrangian for the 2D plasmons can be written as~\cite{sun2018third,Sun_plasmon.2022}:
\begin{align}	\label{eq:grapheneL}
	L &= \sum_{\mathbf{q}} \left( \frac{1}{4\pi qc^2 } \dot{A}_{\mathbf{q}} \dot{A}_{-\mathbf{q}} - \frac{D}{2\pi c^2} A_{\mathbf{q}} A_{-\mathbf{q}} \right) \notag \\
	&\quad + \int \mathrm{d}^2 \mathbf{r} \, \frac{D^{(3)}}{4 c^4} \sum_{i l m n} \Delta_{i l m n} A^i A^l A^m A^n,
\end{align}
where $\mathbf{A}$ is the in-plane component of the dynamical vector potential on the 2D plane and $\mathbf{q}$ the in-plane momentum of the plasmon mode.
Thanks to the simple form of the third-order susceptibility in \equa{eq:chi3}, the corresponding term (plasmon-plasmon interaction term) in \equa{eq:grapheneL} can be expressed using a plain Lagrangian that is local in time.
Restricting to the case where all fields are polarized along the $x$-axis and including a pump field $A_{\text{p}}$ with zero in-plane momentum.
The equation of motion implied by  \equa{eq:grapheneL} for the $\mathbf{q}$-mode is
\begin{align}\label{eqn:parametric_oscillator}
	\left[	\frac{1}{2\pi qc} \partial_t^2 +\frac{\tilde{D}}{\pi c}
	-\alpha A_{\text{p}}^2 \left(1+2\cos(2\omega_{\text{p}}t)\right)
	\right]
	A_{\mathbf{q}}=0
\end{align}
where $\alpha = \frac{9D^{(3)}}{2c^3}$.
Apparently, the pump has created a parametric driving term for this mode, which modulates the mode eigenfrequency periodically at the frequency $2\omega_{\text{p}}$.
Standard solutions of the parametric oscillator give the results in \equa{eq:floquetplasmon_freq}~\cite{Sun_plasmon.2022}.
Furthermore, \equa{eqn:parametric_oscillator} could be numerically diagonalized in the frequency space whose result gives the exact Floquet plasmon bands, corresponding to adding all higher order diagrams beyond \fig{fig:theoryFig}(d)(e).
The $R_{\text{p}}$ in \fig{fig:GrPlasmon}(c) is computed in this way by truncating the replica index with a cutoff $n=\pm 5$.

\subsection{Discussion and experiment protocols for Floquet plasmons}
This section has focused on linearly polarized pump (along $\hat{x}$),
which makes the effective 2D susceptibility $\chi_{\text{eff},ij}$ anisotropic, and 
has focused on the plasmonic momentum $\mathbf{q}$ parallel to the pump field.
The Floquet plasmon dispersion is actually weakly anisotropic on the 2D plane.
For $\mathbf{q}$ along $\hat{y}$, the effective susceptibility $\chi_{\text{eff},yy}$  is still in the form of \equa{eq:grchieff} but with the pump-strength parameter reduced from $3\xi^2/16$ to $\xi^2/16$, which can be inferred from the tensor structure of $\chi^{(3)}$  \equa{eq:chi3}.
For circularly  polarized pump, Floquet plasmon dispersion is isotropic.
Interestingly, the effective susceptibility $\chi_{\text{eff},ij}$ develops a nontrivial Hall response, see \append{appendix:circular}.
The resulting topological properties of the Floquet plasmons deserve further research. 

Note that pristine graphene preserves the spatial inversion symmetry so that the second-order nonlinear susceptibility $\chi^{(2)}$ vanishes at zero momentum. 
However, $\chi^{(2)}$  is nonzero at the nonzero momenta of the plasmons.
This would be an additional contribution to  the effective susceptibility via \fig{fig:theoryFig}(c).
In  \append{appendix:chi2}, we  show that this contribution does not lead to nontrivial band hybridization effects other than simple replicas.

We have focused on the effect of third-order nonlinearity $\chi^{(3)}$ from electronic intraband effects in graphene, see \equa{eq:chi3}. 
In principle, higher order nonlinearities such as $\chi^{(2n+1)}$~\cite{sun2018third} with $n>1$ also contribute to the Floquet effects. 
In graphene, they would lead to similar momentum gaps at band crossing points at $\omega=n\omega_{\text{p}}$.
Their sizes are suppressed by the small parameter $\xi^{2n}$ though.
Furthermore, at higher pump frequencies which are not small compared to the fermi energy, interband transitions have substantial contributions to the optical nonlinearity, whose effects deserve future research.
A similar plasmonic band to \fig{fig:GrPlasmon}(c) was predicted previously~\cite{kiselev2024inducing} with Floquet driving coming from the periodically modulated amplitude of the optical pump, which needs carefully engineered pump lasers.
The effects there involves electronic interband transitions.
We go beyond previous study by demonstrating that  these Floquet plasmonic features arise naturally from a simple infrared pump with mild pump field via intraband optical nonlinearity, while a periodically modulated pump amplitude is not necessary.

To measure these Floquet plasmons  in near-field pump-probe experiments~\cite{Wagner2014UltrafastNanoscaleGraphenePlasmons, Ni2016UltrafastGraphenePlasmons,Fu2024HyperbolicTransientPlasmonsBP}, one should use the mid- to far-infrared pump pulses such as those in \fig{fig:GrPlasmon}(c), and probe pulses at roughly the same frequency and overlapping with the pump pulse in time. 
One should strictly look at the linear response near field signal to the probe pulse, which are the  linear responses of the Floquet plasmons.
Ideally, the unstable plasmons in the flat band in \fig{fig:GrPlasmon}(c) would appear as growing near field signal as the tip moves away from a launcher or an edge of the sample.
For a more quantitative analysis, the real space plasmonic fringes could be Fourier transformed to yield the $R_{\text{p}}$ in \fig{fig:GrPlasmon}(c), which would show the band structure given enough resolution.


\section{\label{sec:phonon}Floquet phonon polaritons in  \lowercase{h}BN}
As a second example, we consider infrared phonon polaritons in 
hexagonal boron nitride (hBN)
~\cite{dai2014tunable,
	Caldwell2014SubDiffractionalHBNPolaritons,	
	Caldwell2015SurfacePhononPolaritons, Li2015HBNHyperbolicPolaritons, 
	Giles2018UltralowLoss, 
	Li2018InfraredHyperbolicMetasurface,
	Ni2021LongLived, 
	Guddala2021TopologicalPhononPolariton,
	Kurman2021SpatiotemporalImaging,
	Guo2025HyperbolicElectroluminescence}
which feature hyperbolic dispersion, low loss and high confinement.
Experiments and theories have shown strong optical nonlinearities in hBN~\cite{Li2013FewLayerMoS2hBN_SHG,Kim2013hBNBilayer_SHG,Yao2021BoronNitrideTwist_SHG,Mueller2025hBN_SFG, ginsberg2023phonon,kang2021ultrafast}.
We will derive the Floquet phonon polaritons in both hBN monolayers  and  thin flakes pumped by infrared lasers.

\subsection{Monolayer hBN}
\begin{figure}
\includegraphics[width=\linewidth]{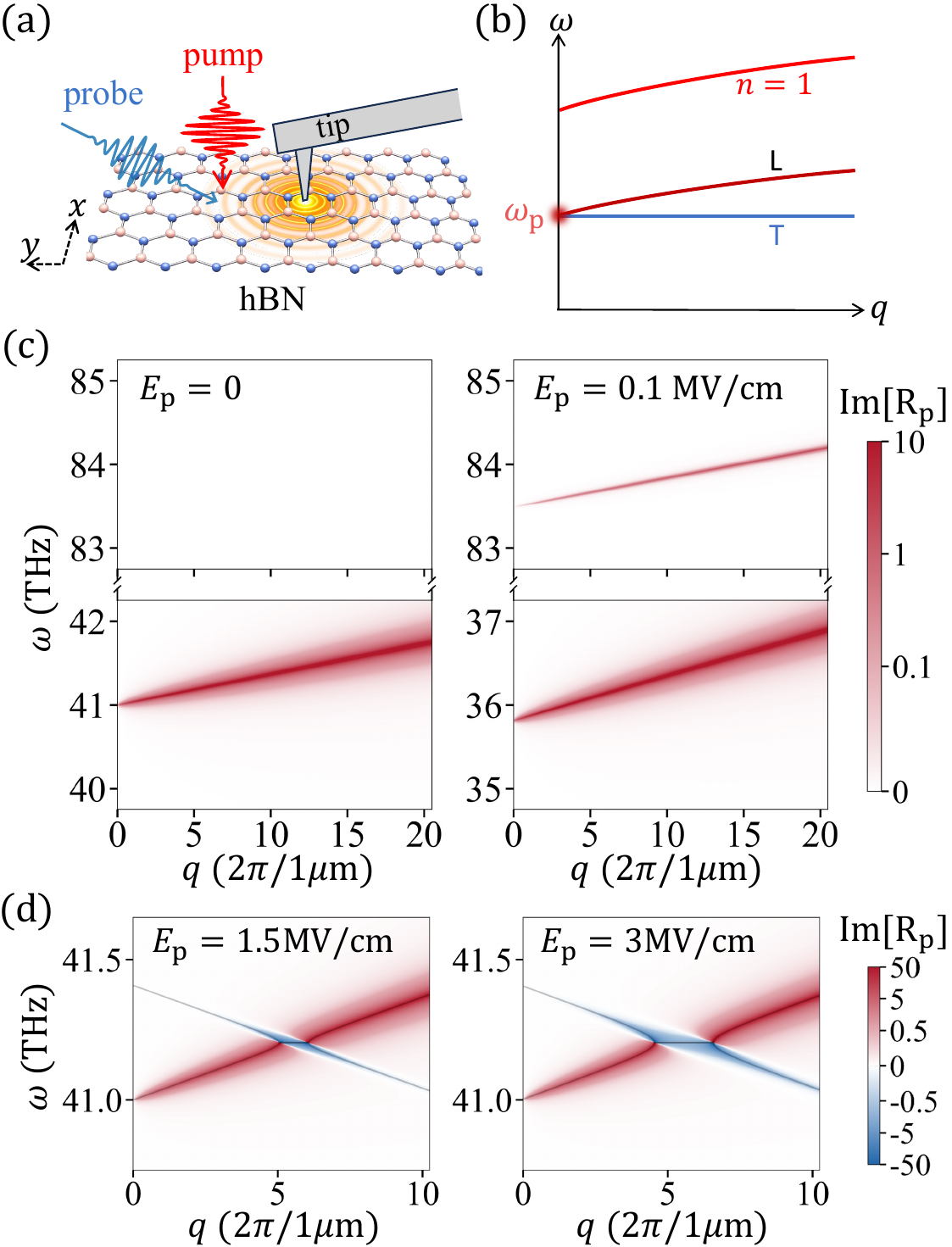}
\caption{\label{fig:hBNFig} 
Floquet phonon polaritons in monolayer hBN.
(a) Schematic of the near-field  pump-probe measurement of Floquet phonon polaritons in monolayer hBN. 
(b) Schematic dispersion of the Floquet phonon polaritons.
(c) Near-field reflection coefficient $R_{\text{p}}$ of monolayer hBN on the frequency-momentum plane computed from \equa{eq:chieffhbn} in equilibrium (left) and with a pump field $E_{\mathrm{p}}=0.1\unit{MV/cm}$ at frequency $\omega_{\mathrm{p}}=\omega_{\mathrm{TO}}=41\unit{THz}$ (right).
In the right panel, the pump induces a new polariton branch around $83.5 \unit{THz}$, and also leads to a red shift of the principal branch.
(d) Same as (c) but with the pump frequency
$\omega_{\mathrm{p}}=2.01 \omega_{\mathrm{TO}}=82.41 \unit{THz}$ at the pump field $E_{\mathrm{p}}=1.5 \unit{MV/cm}$ (left) and  $E_{\mathrm{p}}=3 \unit{MV/cm}$ (right). 
The black curves are real parts of the eigenfrequencies of Floquet polaritons.
Polaritonic flat band appears around the crossing point due to  parametric instability.
For all the plots, the damping rate is $\gamma= 0.1 \unit{THz}$ motivated by that in isotopically pure hBN~\cite{Giles2018UltralowLoss,Ni2021LongLived} and the nonlinear coupling parameters are from Ref.~\cite{iyikanat2021nonlinear}.
}
\end{figure}

Monolayer hBN exhibits 2D phonon polaritons~\cite{Dai2019MonolayerhBNPolaritons,Li2021SuspendedMonolayerhBNPolaritons} as shown in Fig.~\ref{fig:hBNFig}(a).
It also has  second-order nonlinear optical response due to its intrinsically broken inversion symmetry. 
Recent experiments  have already revealed strong second order nonlinear optical effects on monolayer and  few-layer hBN~\cite{Li2013FewLayerMoS2hBN_SHG,Kim2013hBNBilayer_SHG,Yao2021BoronNitrideTwist_SHG,Mueller2025hBN_SFG}.

We begin with a continuous Lagrangian for the dipolar active phonons in monolayer hBN and its coupling to photons:
\begin{align}
	L = &\int \mathrm{d}^2 \mathbf{r}
	\bigg[
	 \frac{1}{2} 
	\left( \dot{\mathbf{u}}^2-\omega_{\text{TO}}^2 \mathbf{u}^2 
	\right)
	+ \lambda_0  \mathbf{u} \cdot \mathbf{E}
	 \notag \\
	&+ \lambda_1  u_x (u_{x}^2-3u_y^2)
	\bigg]
	+ L_{\mathrm{EM}}.
	\label{eq:hBNlagrangian}
\end{align}
Here $\mathbf{u}$ is the (properly normalized) in-plane displacement of the dipolar active mode (TO phonon) with $u_{x/y}$  denoting its $x/y$ components.
Its eigen frequency, $\omega_{\text{TO}}$, is well approximated by a constant one before hybridization with light.	
The second term is its coupling to electric field with the dipole strength $\lambda_0$.
The third term is a cubic nonlinear term which is allowed because of the absence of inversion symmetry, and is constrained to be the form in \equa{eq:hBNlagrangian} by the $C_3$ rotation symmetry and the mirror symmetry under $y \rightarrow -y$, see the coordinate system in \fig{fig:hBNFig}(a).
The parameters $\lambda_0$ and $\lambda_1$ can be determined from experiments and DFT calculations~\cite{ginsberg2023phonon, iyikanat2021nonlinear}.
The $L_{\mathrm{EM}}$ is the Lagrangian of free space photons.
In the near field limit, it reduces to $\sum_{\mathbf{q}} E_{\mathbf{q}} E_{-\mathbf{q}}/(4\pi q)$, the field energy of the longitudinal electic field where $\mathbf{q}$ is the in-plane wave vector.

The intrinsic linear susceptibility thus reads
\begin{align}\label{eq:chi0}
\chi(\omega,\mathbf{q}) 
	= \frac{\lambda_0^2}{\omega_{\text{TO}}^2 - \omega^2}
	\,.
\end{align}
Adding a damping term $i \gamma \omega$ to the denominator and making use of the relation $\epsilon_{\text{3D}}= 1+ 4\pi \frac{1}{a_0} \chi(\omega,\mathbf{q})$ where $a_0$ is the interlayer distance of bulk hBN, one obtains the Lorentzian model for the in-plane dielectric of bulk hBN.
Therefore, the value of $\lambda_0^2$ is set by the experimentally measured dielectric.	
The bare propagator for the longitudinal 2D polariton  is thus found from \equa{eq:G_0nearfield} as
\begin{align}\label{eq:phonon_polariton}
{G}_0^{\text{pl}}(\omega, \mathbf{q}) = 
-2\pi q
\frac{\omega_{\text{TO}}^2 - \omega^2}{\omega_{\mathbf{q}, L}^2 - \omega^2}
\end{align}
in the near field ($q \gg \omega/c$) regime. Here $\omega_{\mathbf{q}, L}=\sqrt{\omega_{\text{TO}}^2+2\pi q \lambda_0^2}$ is the longitudinal polariton frequency which features linear dispersion at low momenta.
In the near field regime, the transverse mode couples very weakly with light and stays as a flat band at $\omega_{\text{TO}}$. 
See \fig{fig:hBNFig}(b) for typical TO-LO splitting in 2D.

We now proceed to the Floquet polaritons in optically pumped hBN monolayer.
For simplicity, we focus on the case where both the (uniform) pump field $E_{\text{p}}e^{-i\omega_{\text{p}}t}\hat{x}+c.c.$ 
and the (polaritonic) probe field
$E_{\text{pr}}e^{i(\mathbf{q}\cdot\mathbf{r}-\omega t)}\hat{x}+c.c.$
are linearly polarized along the x-axis.
In the near field regime where only longitudinal modes are important, this  means that the momentum $\vec{q}$ is also along $\hat{x}$.

In this direction, the cubic ($\lambda_1$) term in \equa{eq:hBNlagrangian} mediates a second order nonlinear susceptibility
\begin{align}\label{eq:chi2}
\chi^{(2)}(\omega_1, \omega_2) 
&= \frac{3 \lambda_1}{\lambda_0^3} 
\chi(\omega_1) \chi(\omega_2) \chi(\omega_1+\omega_2)  
\,
\end{align}
where $\chi(\omega)$ is from \equa{eq:chi0}.
We note again that the linear and nonlinear susceptibilities are defined as the polarization in response to the \textit{net} electric field, so that they have contributions from only 1PI diagrams with respect to photon propagators.
Therefore, the $\chi$ in \equa{eq:chi2} should not include polaritonic effects, but are bare phonon propagators.
By acting two times, the cubic term also mediates a third-order nonlinear susceptibility:
$
	\chi^{(3)}(\omega_1, \omega_2, \omega_3) 
	= \frac{9 \lambda_1^2}{\lambda_0^6} 
	\chi(\omega_1) \chi(\omega_2) \chi(\omega_1+\omega_2)  
\chi(\omega_3)  \chi(\omega_1+\omega_2+\omega_3)  
$,
and by induction, even higher order ones.
Obviously, from \fig{fig:theoryFig}(c)(d), these nonlinear coefficients will contribute to the effective linear susceptibility.
For example, the diagram in  \fig{fig:theoryFig}(c) contains the information of $n=-1$ (and $n=+1$ if one changes $\omega_{\text{p}}$ to $-\omega_{\text{p}}$) replica band.
Actually, a re-summation of  infinite diagrams can be done which gives the effective susceptibility 
\begin{align}\label{eq:chieffhbn}
\chi_{\text{eff}}(\omega,q) 
=
&\lambda_0^2 
\bigg/
\bigg[
\omega_{\text{TO}}^2 - \omega^2-
\notag\\
& 
E_{\text{p}}^2 
	\frac{36 \lambda_1^2  \lambda_0^2}{(\omega_{\text{TO}}^2 - \omega_{\text{p}}^2)^2}
	G_0^{\text{phonon}}(\omega \pm \omega_{\text{p}},q)
\bigg]
\,,
\end{align}
see \fig{fig:SIFig2} and \append{appendix:hBN} for a derivation.
Here we have defined the 
polariton propagator similar to \equa{eq:phonon_polariton} but defined in terms of the phononic response function:   $G_0^{\text{phonon}}(\omega,\mathbf{q}) 
= 1/(\omega_{\mathbf{q},L}^2 - \omega^2)$.
With a nonzero phononic damping rate $\gamma$, one needs to change $\omega^2$ to $\omega (\omega+i\gamma)$, $\omega_{\text{p}}^2$ to $\omega_{\text{p}} (\omega_{\text{p}}+i\gamma)$, and 
$(\omega\pm\omega_{\text{p}})^2$ to $(\omega\pm\omega_{\text{p}}) (\omega\pm\omega_{\text{p}}+i\gamma)$
in the propagators.
This set of diagrams exhausts the Floquet effects with the cutoff $\pm 1$ in the Floquet replica.  
The physical meaning is that 
the probe field $E_{\text{pr}}(\omega,q)$ with a pump field $E_{\text{p}}(\omega_{\text{P}})$ induces a phononic displacement $u(\omega+\omega_{\text{P}},q)$ through the cubic term in \equa{eq:hBNlagrangian}, which propagates as the polariton propagator $G_0^{\text{phonon}}(\omega,\mathbf{q})$.
Afterwards, the phononic displacement combines with a negative-frequency  pump field 	$E_{\text{p}}(-\omega_{\text{P}})$ to induce the polarization $P(\omega,q)$.
Summing all repetitions of this process gives the  effective linear susceptibility of pumped monolayer hBN in \equa{eq:chieffhbn} seen by the polaritonic field. 

The Floquet polariton dispersion is set by the zeros of $\epsilon_{\mathrm{2D}}=1+2\pi q\chi_{\text{eff}}$. 
Substituting \equa{eq:chieffhbn} into this condition, one finds solutions near the original polariton
$\omega_{\mathbf{q}, L}$ and 
additional solutions near $\omega_{\mathbf{q}, L}\pm\omega_{\mathrm{p}}$, which are Floquet phonon-polariton sidebands. 
Fig.~\ref{fig:hBNFig}(c) shows 
the near-field optical reflection coefficient 
$R_{\text{p}}= 1- 1/\epsilon_{\text{2D}}$
of monolayer hBN using the first-principle parameters $\lambda_0, \lambda_1$ from Ref.~\cite{iyikanat2021nonlinear}.
Upon pumping, the emergence of a new Floquet polariton branch dispersing from $\omega_{\text{TO}}+\omega_{\mathrm{p}}$ is clearly visible. 
This effect is strongest when $\omega_{\mathrm{p}}\approx\omega_{\text{TO}}$ so that the TO phonon is resonantly pumped. 
In this case, the emergence of this new mode may also be understood as a pump-enhanced parametric down conversion process: an incident photon at $2\omega_{\text{TO}}$ can resonantly down convert into two phonons at $\omega_{\text{TO}}$.
Note that this down conversion channel exists even  in equilibrium through $\chi^{(2)}$ and the quantum/thermal fluctuations of the TO and LO phonons.
However, because of the dispersion of the longitudinal phonons, this channel only contributes a very 	weak and broad absorption peak around $2\omega_{\text{TO}}$, which can hardly be viewed as a collective mode. 
Upon resonant pumping of the zero-momentum TO phonon, the down conversion channel of $2\omega_{\text{TO}}$ photons into this mode is greatly enhanced, leading to a sharp absorption peak at $2\omega_{\text{TO}}$.
Therefore, the new polariton branch  at $2\omega_{\text{TO}}$ exists even after a  pulsed pump is gone while the TO phonon is still oscillating within its lifetime.

Through $\chi^{(2)}$, the polaritons of the principal branch are also parametrically driven by the pump itself, with parametric resonance realized when the pump is at a frequency slightly above $2\omega_{\text{TO}}$, see Fig.~\ref{fig:hBNFig}(d).
There the polariton branch around $-\omega_{\text{TO}}$ is upshifted by the pump to give a replica with negative spectral weight, similar to the case of Floquet graphene plasmons.
This replica crosses the principle branch and leads to flat bands with unstable Floquet polaritons, exhibiting the same physical phenomena as those of Floquet graphene plasmons in \fig{fig:GrPlasmon}.


\subsection{hBN  flakes}
\label{sec:hBN_flake}

\begin{figure}
	\includegraphics[width=\linewidth]{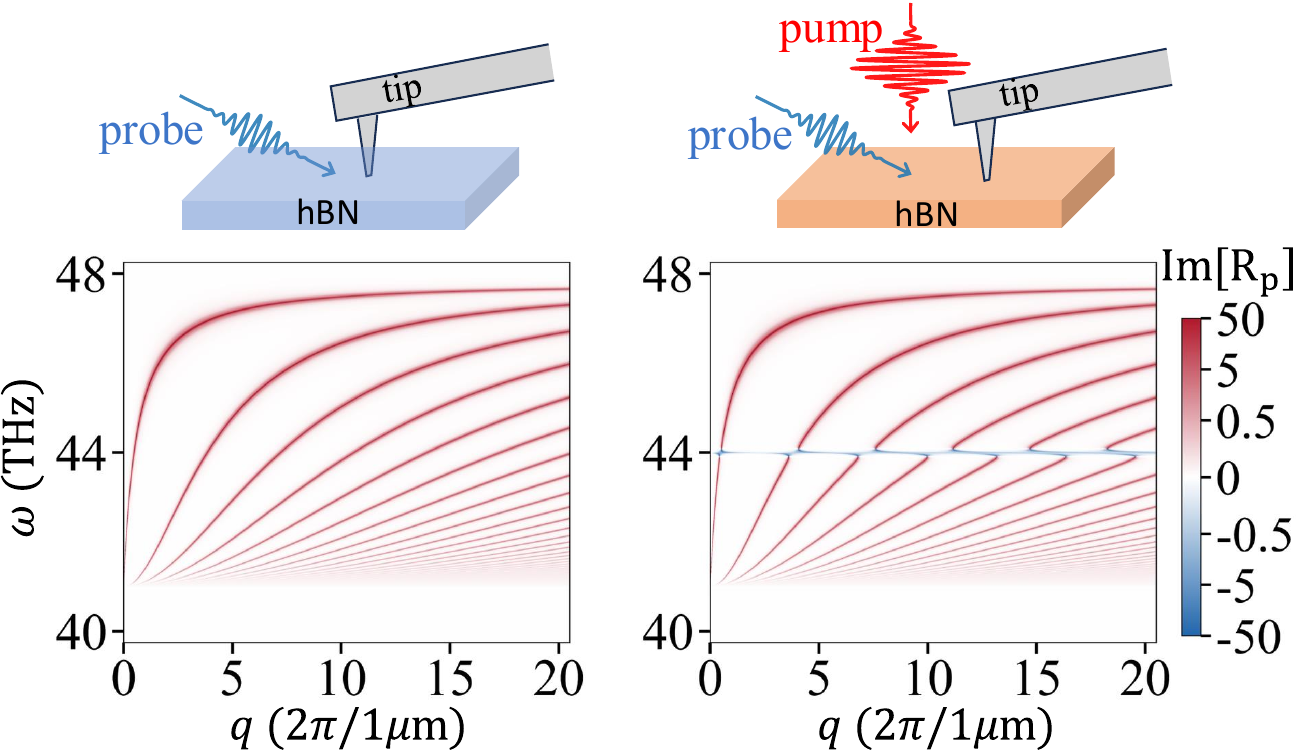}
	\caption{\label{fig:hBN_flake} 
		Floquet phonon polaritons in hBN thin flakes. 
		The left panel is the near field reflection coefficient $\mathrm{Im}[R_{\text{p}}]$ of an equilibrium hBN thin flake plotted on the frequency-momentum plane.
		The right panel is the same but under uniform pump field with the frequency $\omega_{\mathrm{p}}=2.07 \omega_{\mathrm{TO}}=85 \unit{THz}$ and field strength $E_{\mathrm{p}}=8 \unit{MV/cm}$, within the reach of modern table top lasers~\cite{Wang2026FloquetGapGraphene}.
		The parameters used include the flake thickness $d=100 \unit{nm}$, the phononic damping rate $\gamma= 0.1 \unit{THz}$~\cite{Giles2018UltralowLoss,Ni2021LongLived},
		the interlayer distance $a_0=0.33 \unit{nm}$, the  high frequency asymptotic in-plane dielectric $\epsilon_{\perp, \infty}=4.87$.
		The out-of-plane dielectric is set as $\epsilon_{z}=3.0$ which is a good approximation to hBN in this frequency range.	
		The pump light has the wavelength $1.6 \unit{\mu m}$ inside the flake, much larger than the flake thickness so that it is well approximated by a uniform field.
	}
\end{figure}

For a thin hBN flake that contains $N \gg 1$ layers, the $\chi^{(2)}$ responses between adjacent layers cancel each other in the continuous limit, so that the lowest order optical nonlinearity is a $\chi^{(3)}$ nonlinearity for uniform electric field.
This is consistent with the restoration of inversion symmetry of bulk hBN.
Specifically, because two adjacent layers of hBN are rotated by $180$ degrees with respect to each other, the $\lambda_1$ in \equa{eq:hBNlagrangian} has alternating signs among layers.
Summing all layers, each contributing an \equa{eq:chi2},  the total $\chi^{(2)}$ vanishes in the continuous limit.
As a result, only odd-order optical nonlinearities contribute to the pump induced effective susceptibility, and the bulk effective susceptibility reduces to
\equa{eq:chieffhbn} with the polariton propagator $G_0^{\text{phonon}}$ replaced by the bare phonon propagator $G_{\text{T}}^{\text{phonon}}$ defined as  $G_{\text{T}}^{\text{phonon}}(\omega,\mathbf{q}) 
= 1/[{m(\omega_{\text{TO}}^2 - \omega^2)}]$, 
see \append{appendix:hBN} for a derivation.
The corresponding 3D effective in-plane dielectric is 
\begin{align}\label{eq:chieffhbn_3D}
	\epsilon_{\perp, \text{eff}}(\omega,q) 
	=& \epsilon_{\perp, \infty} + 4\pi
	\frac{\lambda_0^2}{a_0}
	\bigg/
	\bigg[
	\omega_{\text{TO}}^2 - \omega^2-
	\notag\\
	& 
	E_{\text{p}}^2 
	\frac{36 \lambda_1^2  \lambda_0^2}{(\omega_{\text{TO}}^2 - \omega_{\text{p}}^2)^2}
	G_{\text{T}}^{\text{phonon}}(\omega \pm \omega_{\text{p}},q)
	\bigg]
	\,,
\end{align}
where $a_0$ is the interlayer distance and dielectric constant $\epsilon_{\perp, \infty}$ is the high frequency asymptotic value of the dieletric arising from high-energy electronic transitions beyond the low-energy phonon dynamics in our model.
The rule of adding the damping rate is the same as that for \equa{eq:chieffhbn}.

The lowest order term in \equa{eq:chieffhbn_3D} contributed by the pump is understood as the second diagram in \fig{fig:SIFig2}(b).	
The physical picture is that the probe field and the  uniform pump field induce a second order phononic displacement $u(\omega+\omega_{\text{P}},q)$  via the second order phononic nonlinearity (the $\lambda_1$ cubic term) in \equa{eq:hBNlagrangian}.
Contrary to the monolayer case, this phononic displacement has alternating signs among layers in the bulk, which has zero dipole and does not emit EM field and thus does not have polaritonic corrections.
In other words, the mode excited via the second order process is one with a large z-direction momentum that is atomic scale, which is a neutral mode whose eigen frequency is simply $\omega_{\text{TO}}$.
After propagation with $G_{\text{T}}^{\text{phonon}}(\omega + \omega_{\text{p}},q)$, this phononic displacement combines with a negative-frequency  pump field 	$E_{\text{p}}(-\omega_{\text{P}})$ to induce the displacement $u(\omega,q)$ via a second application of the $\lambda_1$  term.
This displacement, at order $\lambda_1^2$,  has the same sign among layers now so that it has an electrical polarization, thus contributing an effective susceptibility in \equa{eq:chieffhbn_3D}.

A typical Floquet effect of the hyperbolic phonon polaritons in pumped hBN is shown in \fig{fig:hBN_flake}.
Again, what is plotted is the imaginary part of the near field reflection coefficient $\mathrm{Im}[R_{\text{p}}]$ of an hBN flake, see, e.g., Eq.~B8 in Ref.~\cite{sun2020collective} for its expression and derivation.
In the right panel of \fig{fig:hBN_flake}, there is a $-\omega_{\text{TO}}+\omega_{\mathrm{p}}$ replica band at about $44 \unit{THz}$ which comes from shifting the neutral phonon mode at $-\omega_{\text{TO}}$ up by $\omega_{\mathrm{p}}$.
Therefore, it has negative spectral weight as shown by its blue color, following the same principle 	discussed before.
This replica crosses the original hyperbolic polariton branches to give polaritonic flat bands with unstable modes and exceptional points, similar to \fig{fig:GrPlasmon}(c) and \fig{fig:hBNFig}(d).
The physical reason for these unstable modes is a parametric instability from the decay of the pump photon to a hyperbolic polariton and a neutral phonon.

Experimentally, for both hBN monolayer and 	flakes, the new poles around $\omega_{\text{TO}}+\omega_{\mathrm{p}}$ in the effective susceptibility in \equa{eq:chieffhbn} and
around $-\omega_{\text{TO}}+\omega_{\mathrm{p}}$ in \equa{eq:chieffhbn_3D} can be simply measured by far-field optical pump-probe technique.
To measure the dispersions of Floquet phonon polaritons in Fig.~\ref{fig:hBNFig}(c)(d) and \fig{fig:hBN_flake}, the near-field pump-probe setup~\cite{Wagner2014UltrafastNanoscaleGraphenePlasmons, Ni2016UltrafastGraphenePlasmons,Fu2024HyperbolicTransientPlasmonsBP} may be used, as shown schematically in Fig.~\ref{fig:hBNFig}(a) and \fig{fig:hBN_flake}.

\section{\label{sec:jplasmon}Floquet Josephson plasmons in layered superconductors}
\begin{figure}
	\includegraphics[width=0.85 \linewidth]{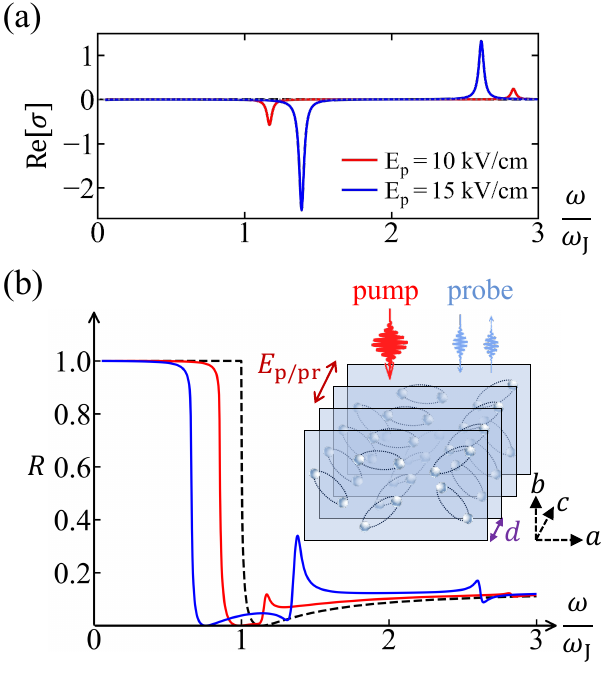}
	\caption{\label{fig:JP} 
		Floquet Josephson plasmons in layered superconductors.
		(a) Real part of the optical conductivity $\sigma=-i\omega(\epsilon_{\text{eff}}-1)/(4\pi)$ in units of $\epsilon_\infty\omega_{\mathrm{J}}/4\pi$ along the out-of-plane direction of a pumped layered superconductor plotted from \equa{eq:JPeeff}. 
		The black dashed curve corresponds to equilibrium.
		The red (blue) solid curve is that under the pump field $E_{\mathrm{p}}=10\unit{kV/cm}$ ($15\unit{kV/cm}$) with the frequency $\omega_{\mathrm{p}}=\omega_{\mathrm{J}}$. 
		The peaks at nonzero frequencies are  induced by the pump.
		(b) The corresponding optical reflectivity $R=|(1-\sqrt{\epsilon_{\text{eff}}})/(1+\sqrt{\epsilon_{\text{eff}}})|^2$ showing pump-induced Floquet Josephson plasmon resonances. 
		The inset shows the experimental geometry for the layered superconductor with the pump and probe electric fields polarized along the c-axis.
		The other parameters used are $\omega_{\text{J}}=0.5\unit{THz}$,  $\epsilon_\infty=4.5$, and $d=0.6\unit{nm}$ motivated by those of $\mathrm{La}_{2-x}\mathrm{Ba}_{x}\mathrm{CuO}_{4}$~\cite{zhang2023revealing}. 
		We have added a nonzero damping rate $\gamma=0.02 \unit{THz}$ to the Josephson plasmons by replacing $(\omega \pm 2k\omega_{\text{p}})^2$ by $(\omega \pm 2k\omega_{\text{p}})(\omega \pm 2k\omega_{\text{p}}+i\gamma)$ in \equa{eq:JPeeff}, which could be caused by thermally excited quasi-particles and pair-excitations.
	}
\end{figure}

As the last example, we consider Josephson plasmons~\cite{sun2020collective, Sellati2023GeneralizedJosephsonPlasmons,Fiore2024InvestigatingJosephsonPlasmons} in layered superconductors such as cuprates where superconducting ab-planes are coupled via Josephson tunneling along the c-axis~\cite{Basov2005ElectrodynamicsHighTc, Sellati2023GeneralizedJosephsonPlasmons,Fiore2024InvestigatingJosephsonPlasmons}, as shown in Fig.~\ref{fig:JP}(b). 
In these systems, inter-layer Josephson coupling leads to giant third-order susceptibility $\chi^{(3)}$ in the terahertz range along the c-axis~\cite{Rajasekaran2016JosephsonAmplification,Gabriele2021NonlinearTHzCuprates,kaj2023terahertz,zhang2023revealing}.
Recent experiments have  revealed pronounced nonlinear signatures near the Josephson plasma edge~\cite{nicoletti2022coherent,zhang2023revealing}. 
Therefore, under a strong THz pump along the c-axis, we expect the excitations to be modified to Floquet Josephson plasmons.
For example, the process in \fig{fig:theoryFig}(e) leads to Floquet  plasmon replicas at $\omega_{\text{J}} \pm 2\omega_{\text{p}}$ where $\omega_{\text{J}}$ is the original Josephson plasma frequency.

To simplify our analysis, we focus on the case that both the pump and probe pulses are normally incident on the ac-plane with their electric field polarized along the c-axis, which is a typical far field experimental configuration as shown in Fig.~\ref{fig:JP}(b). 
We further assume that the penetration depth of the pump is much larger than that of the probe, so that the pump field is effectively a uniform one in our model.
In the gauge of zero electric potential so that $\mathbf{E}=-\partial_t \mathbf{A}/c$,
the vector potential of the pump/probe field is $\mathbf{A}_{\text{p}/\text{pr}}(t)=A_{\text{p}/\text{pr}}(t)\hat{c}$. 
In the long wavelength limit, the three dimensional (3D) Lagrangian density of the superconducting order parameter phase $\theta$ and the EM field reads~\cite{sun2020collective}
\begin{equation}\label{eq:Lforlayered_0}
\mathcal{L}=
	\frac{\nu}{2} \dot{\theta}^2
	+E_{\text{J}}
	\cos \left[d \left(\partial_z\theta + \frac{2e}{\hbar c}A\right) \right]
	+\mathcal{L}_{\text{EM}}[A]
,
\end{equation}
where $\nu$ is a 3D density of states, $E_{\text{J}}$ is a 3D density of Josephson coupling energy and $d$ is the interlayer distance.
We have focused on the case where there is only one superconducting layer in a unit cell such as the case of 
$\mathrm{La}_{2-x}\mathrm{Ba}_{x}\mathrm{CuO}_{4}$~\cite{zhang2023revealing}. The extension to `bilayer' superconductors such as  $\mathrm{YBa}_{2}\mathrm{Cu}_{3}\mathrm{O}_{7-\delta}$ is straightforward.
The free Lagrangian density for the EM field is 
$
\mathcal{L}_{\text{EM}}[\mathbf{A}]=
\left[(\partial_t \vec{A}/c)^2 
- (\nabla \times \vec{A})^2
\right]/(8\pi)
$.

We just need to derive the current response to the pump and probe fields in the uniform limit.
Solving the equation of motion for $\theta$ from \equa{eq:Lforlayered_0}, one finds that it does not respond to the uniform perturbing field.
Therefore, one may just set $\theta=0$ so that the Lagrangian simplifies to the following form:
\begin{equation}\label{eq:Lforlayered}
	\mathcal{L}=
\frac{1}{8\pi c^2}(\partial_t A)^2
	+E_{\text{J}}
	\cos \left(\frac{2e d}{\hbar c}A \right)
\,.
\end{equation}
From Eq.~(\ref{eq:Lforlayered}), one obtains the Josephson current density  and the equation of motion for the vector potential along the c-axis: 
\begin{align}
& j  = -\frac{2e d E_{\text{J}}}{\hbar} \sin\left(\frac{2e d}{\hbar c} (A_{\text{pr}}+A_{\text{p}}) \right), 
\label{jp1}
\\
& \partial_t^2 A_{\text{pr}}	 = 4\pi c j
\label{AandJ}
\,.
\end{align}
The pump field $A_{\text{p}}$ enters the expression of the Josephson current. However, it does not enter \equa{AandJ} because it is defined as the total incident field from the pump laser that has penetrated into the sample, while any feedback effect from the sample itself has already been included in $A_{\text{p}}$.
Solving Eqs.~\eqref{jp1} \eqref{AandJ} with the linear order expansion of the $\sin$ term gives the Josephson plasmon oscillation solution $A_{\text{pr}} \propto e^{\pm i\omega_{\text{J}} t}$ with the Josephson plasma frequency 
$\omega_{\text{J}}=4ed \sqrt{\pi E_{\text{J}}} /\hbar$.
Indeed, \equa{jp1} implies that  the c-axis linear optical conductivity is 
$\sigma=i\omega_{\text{J}}^2/(4\pi \omega)$, corresponding to the
susceptibility 
$
\chi=-\omega_{\text{J}}^2/(4\pi \omega^2)
$
and the dielectric $\epsilon=1-\omega_{\text{J}}^2/\omega^2$.

Expanding the $\sin$ term in \equa{jp1} to the third-order, one obtains the third-order susceptibility $\chi^{(3)}(\omega_1,\omega_2,\omega_3)=- (\frac{2e d}{\hbar })^2  \omega_{\text{J}}^2/(24\pi \omega_s \omega_1 \omega_2 \omega_3)$ where $\omega_s=\omega_1 + \omega_2 + \omega_3$.
By computing the diagram in \fig{fig:theoryFig}(e) under the pump field $E_{\text{p}}e^{-i\omega_{\text{p}}t}\hat{c}+c.c.$, 
one obtains the effective linear susceptibility that contains additional poles at $\pm \omega_{\text{J}} \pm 2\omega_{\text{p}}$.
These poles would certainly contribute peaks in the optical conductivity and corresponding features in the reflectivity of the probe, see \fig{fig:JP}. 
Note that because inversion symmetry forbids even-order optical nonlinearities, the pump-induced correction to the linear response can only involve even powers of the pump field.
Consequently, the Floquet replicas of the Josephson plasmon are shifted only by even multiples of $\omega_{\mathrm{p}}$.

In fact, an effective linear susceptibility  that is   non perturbative in the pump field can be obtained.
Noting $A_{\text{p}}=2A_{\text{p}0}\cos(\omega_{\text{p}}t)$, the current  in  Eq.~(\ref{jp1}) is expanded to first order in $A_{\text{pr}}$ as
\begin{align}
	j \approx &-\frac{\omega_{\text{J}}^2}{4\pi c}A_{\text{pr}}(t)
	\bigg[
	J_0(2\varphi_{\text{p}})
	\notag\\
	&+2\sum_{k=1}^{\infty}(-1)^kJ_{2k}(2\varphi_{\text{p}})\cos(2k\omega_{\text{p}}t) 
	\bigg]
	\label{eq:Besselexpansion}
\end{align}
where $J_{n}$ is the $n$-th Bessel function of the first kind, and we have defined the dimensionless pump strength $\varphi_{\text{p}}=\frac{2ed}{\hbar c}A_{\text{p}}$.
Closing \equa{AandJ} with \equa{eq:Besselexpansion}, one observes that the second term in Eq.~(\ref{eq:Besselexpansion}) contains all the periodic driving terms for the Josephson plasmon.
Solving \equa{AandJ} with \equa{eq:Besselexpansion} self consistently with the cutoff $n=\pm 1$ for the Floquet driving terms, one obtains the effective linear susceptibility $\chi_{\text{eff}}$.
The resulting effective dielectric $\epsilon_{\text{eff}}= 1+ 4\pi \chi_{\text{eff}}$ follows as
\begin{align}
\frac{\epsilon_{\text{eff}}}{\epsilon_\infty}
=1-\frac{\omega_{\text{J}}^2}{\omega^2}
\bigg[ 
& J_0(2\varphi_{\text{p}})
-
\sum_{k=1}^{\infty}\frac{J_{\text{2k}}(2\varphi_{\text{p}})^2}{J_0(2\varphi_{\text{p}})-(\omega \pm 2k\omega_{\text{p}})^2/\omega_{\text{J}}^2}
\bigg]
\label{eq:JPeeff}
\end{align}
where we have added the background dielectric constant $\epsilon_\infty$ arising from high-energy electronic and lattice polarizations beyond the low-energy Josephson dynamics.
The route from \equa{eq:Besselexpansion} to \equa{eq:JPeeff} is similar to that from \equa{eq:interpe} to \equa{eq:chieff}, only that the expansion in \equa{eq:Besselexpansion} is not in powers of the pump field but in orders of Bessel functions.
Again, we have added a partial self-energy correction implied by the first term of \equa{eq:Besselexpansion} to the plasmon propagator in \fig{fig:theoryFig}(e), so that its Josephson plasmon frequency is renormalized to $\omega_{\text{J}}\sqrt{J_0(2\varphi_{\text{p}})}$.

In the limit of weak pump field, $\varphi_{\text{p}}\ll 1$, the Bessel functions reduce to $J_0(2\varphi_{\text{p}})\approx 1$ and $J_{2}(2\varphi_{\text{p}})\approx (\varphi_{\text{p}})^{2}/2$. Under this assumption, 
$
\epsilon_{\text{eff}}/\epsilon_\infty\approx1 - \frac{\omega_{\text{J}}^2}{\omega^2}
\left[1-
\frac{\varphi_{\text{p}}^4/4}{1-
(\omega \pm 2\omega_{\text{p}})^2/\omega_{\text{J}}^2}
\right]
$ 
reduces to that from \fig{fig:theoryFig}(d)(e) with $\chi^{(3)}$ derived from Eq.~(\ref{jp1}). 
The Floquet Josephson plasmons are identified as $\epsilon_{\text{eff}}=0$, which occur around $\pm \omega_{\text{J}}\pm 2k\omega_{\text{p}}$ in the weak pump ($\varphi_{\text{p}}\ll 1$) case.

For stronger pumps, the pump-induced  renormalization of the plasmons become apparent.
The optical conductivity $\sigma_{\text{eff}}=(\epsilon_{\text{eff}}-1)\omega/(4\pi i)$ and reflectivity spectra
obtained from Eq.~(\ref{eq:JPeeff}) are shown in Fig.~\ref{fig:JP}, 
where Floquet Josephson plasmons appear close to 
$\pm \sqrt{J_0(2\varphi_{\text{p}})}  \omega_{\text{J}} + 2\omega_{\text{p}}$.
The fundamental mode is red shifted compared to the equilibrium case, which is expected from the concave shape of the $\sin$ function in \equa{jp1}.
Notably, the spectral weight of the Floquet branch at $- \sqrt{J_0(2\varphi_{\text{p}})}  \omega_{\text{J}} + 2\omega_{\text{p}}$ is negative 
because it comes from  shifting the negative frequency mode up by $2\omega_{\text{p}}$ to positive frequency, the same reason as the graphene plasmon case 	in  \equa{eq:grchieff}.
This indicates that the pumped system behaves as a gain medium around that frequency.
In principle, the reflectivity could be larger than unity in certain frequency ranges. 
However, the reflection problem at the interface between vacuum and an infinitely deep gain medium described by \equa{eq:JPeeff} involves subtleties~\cite{Mathey.2022,Yang.2026}, which deserves further study using the Fresnel-Floquet approach~\cite{Michael.2022_Floquet_Fresnel}.
Moreover, we have made the simplifying assumption of uniform pump field so that analytical results may be obtained, while a self-consistent treatment with a finite penetration depth of the pump~\cite{Yang.2026} is left for further study.

Experimentally, this effect can be  measured by far field THz pump-probe technique shown schematically in the inset of Fig.~\ref{fig:JP}(b), such as the 2D coherent  spectroscopy~\cite{zhang2023revealing, gao2023twodimensional,gomez2024principles,Liu2024JosephsonEcho,fiore2025twodimensional}.	
The predictions in Fig.~\ref{fig:JP}(a)(b) are the linear response signals to the probe pulse which overlaps with the pump pulse in time.
Of course, these Floquet Josephson plasmons are also visible to near field probes as sharp peaks in the $R_{\text{p}}$.
In light of recent advances in THz near-field techniques~\cite{
Zhang2018TerahertzNanoimagingGraphene,	
Gallagher2019QuantumCriticalDiracFluid,
	Jing2021THzWTe2, 
	Zhao2023HydrodynamicPlasmonsGraphene,
	Chen2023AnisotropicAcousticTHzPlasmons,
	Xu2024DiracFluidNanoterahertzPolaritons}, this type of experiment appears promising.

\section{Discussion}
We have developed a practical theoretical method for describing Floquet polaritons in optically driven materials.
This is achieved by deriving the effective susceptibilities, Eqs.~\eqref{eq:chieff}, \eqref{eq:grchieff}, \eqref{eq:chieffhbn} and \eqref{eq:JPeeff}, which 
 are conveniently computed from the nonlinear optical susceptibilities of the system. The latter are either known experimentally or can be derived from theory unambiguously.
In this way, the Floquet polariton spectrum is predicted quantitatively without artificial coupling parameters.
Examples in this work include the Floquet graphene plasmons in \fig{fig:GrPlasmon}, Floquet phonon polaritons in \fig{fig:hBNFig}, and Floquet Josephson plasmons in  \fig{fig:JP}.
These predictions can be readily verified by either near field~\cite{Wagner2014UltrafastNanoscaleGraphenePlasmons, Ni2016UltrafastGraphenePlasmons} or far-field optical pump-probe experiments, with the two-dimensional coherent spectroscopy measurement~\cite{zhang2023revealing,gao2023twodimensional,gomez2024principles,fiore2025twodimensional} specifically targeted for this purpose.

Note that in the examples of graphene plasmons and hBN phonon polaritons, to focus on the essential physics, we have neglected the screening from the substrate for simplicity. 
For the 2D samples of interest, the effect of a linear substrate  is to modify the kernel $2\pi q$ in both the numerator and  denominator of \equa{eq:G_0nearfield} to
$2\pi q (1+R_{\text{sub}})$, where $R_{\text{sub}}$ is the near field reflection coefficient at the interface between the system and the substrate.
The other formulas are unchanged.
The polariton spectra is thus modified but the essential physics of Floquet engineering stays the same.
Similarly, for the thin flakes, $R_{\text{sub}}$ modifies the net near field reflection coefficient  of the flake via a simple formula, see, e.g., section~3 in the supplemental information of Ref.~\cite{Ni2021LongLived}.

In the future, it is worthwhile to apply this method to other interesting systems driven by light.
For example, exciton polaritons in either artifical cavities or natural semiconductor flakes have strong $\chi^{(3)}$ nonlinearities~\cite{Carusotto2013QuantumFluids,
	Savvidis2000AngleResonantStimulatedPolaritonAmplifier,
	Saba2001HighTemperatureParametricAmplification,
	Barachati2018InteractingPolaritonFluids,
	Wu2021NonlinearParametricScattering,
	Zhao2022NonlinearPolaritonParametricEmission,
	Xiang2026DipolarPolaritons}.
Under pump light at frequency $\omega_{\text{p}}$, the resulting  Floquet exciton polaritons should exhibit similar flat bands and exceptional points to \fig{fig:GrPlasmon} due to parametric instability.
Indeed,  Floquet driving effects of excitons have been observed before~\cite{Uchida2022CoherentExcitonFloquetStates,Kobayashi2023FloquetExcitons}.
It is also interesting to extend the study of Floquet polaritons to
correlated electron systems driven by light, such as
 superconductors~\cite{Saveliev2010JosephsonPlasmaWaves,
Laplace2016JosephsonPlasmonics, sun2020collective,nicoletti2022coherent,kaj2023terahertz,zhang2023revealing,
	Sellati2023GeneralizedJosephsonPlasmons,Fiore2024InvestigatingJosephsonPlasmons}, excitonic insulators~\cite{Murakami.2020, SunMillis2020BardasisPolaritons, Xuan.2026, shao2026electromagneticresponsesbilayerexcitonic}, magnetic systems~\cite{Huebl2013MagnonPhoton,Tabuchi2014MagnonPhoton},
	fractional quantum Hall systems~\cite{Smolka2014CavityQED2DEG, Ravets2018PolaronPolaritonsQH} and Wigner crystals~\cite{zhao2025electronicphononsmoireelectron, dong2026opticaldetectionmanipulationpseudospin}.


\appendix

\section{Keldysh formalism for Floquet polaritons}
\label{app:keldysh}
The non-equilibrium Green function~\cite{keldysh1964diagram,schwinger1961brownian,altland2010condensed,kamenev2023field} approach  is a rigorous formalism for the Floquet polariton problem, i.e., the problem of a driven-open system which has dissipation and fluctuations.  
We now present the  formulation of non-equilibrium Green functions and its perturbative expansion in the pump field  based on Keldysh path integral~\cite{keldysh1964diagram,schwinger1961brownian,altland2010condensed,kamenev2023field}.	
In addition to the classical mean field saddle point discussed in the main text, this exact formulation also incorporates the effect of thermal and quantum fluctuations.

\subsection{The action for light and matter}
\label{app:action}
The generic Lagrangian describing a material system coupled to the electromagnetic field can be written as
\begin{align}
	L=L_{\text{M}}[\phi,\mathbf{A}]+L_{\text{EM}}[\mathbf{A}] \,,
\end{align}
where
$L_{\text{M}}$ describes the material degrees of freedom $\phi$ and their coupling with the electromagnetic fields $\mathbf{A}$. Integrating out the matter fields $\phi$ yields an effective action $S[\mathbf{A}]$ for $\mathbf{A}$ that is nonlocal both in space and time.
The quadratic terms in $S[\mathbf{A}]$ contain the information of linear optical susceptibility  $\chi$  of the material, giving a propagation kernel for the electric fields $\mathbf{E}$, as shown in Fig.~\ref{fig:theoryFig}(b), which defines the bare polariton propagator, i.e., the bare Green function $G_0(\omega,\mathbf{k})$.
The higher-order terms (beyond $A^2$)  in $S[\mathbf{A}]$  contain the nonlinear optical susceptibilities $\chi^{(i+1)}$. 

In the presence of a strong pump field, the EM field $\mathbf{A}$ in the Lagrangian is replaced by $\mathbf{A}+2\mathbf{A}_{\text{p}}\cos(\omega_{\text{p}}t)$, where the pump field enters as a classical background. 
The problem becomes that of a bosonic system with nonlinear interactions and  periodically driven.
To obtain the information of Floquet polaritons, 
the method we implement is to derive an effective time-translational-invariant action for the polaritonic field $\mathbf{A}$   that incorporates the influence of the pump. This leads to a renormalized propagation kernel for the electric field $\mathbf{E}$, which gives the dressed Green function $G(\omega,\mathbf{k})$  in \equa{eq:dyson}.

\subsection{Keldysh formalism for polaritonic Green functions}
We work in the Weyl gauge $\varphi=0$ so that the electric field is related to the vector potential as $\mathbf{E}=-\partial_t \mathbf{A}/c$. The generating functional on the Keldysh contour $\mathcal{C}$~\cite{keldysh1964diagram,schwinger1961brownian,altland2010condensed,kamenev2023field} is
\begin{align}
Z&=\int \mathcal{D}[\phi_+,\phi_-,\mathbf{A}_+,\mathbf{A}_-]\,e^{iS_{\mathrm{EM}}[\mathbf{A}_+,\mathbf{A}_-]+iS_{\mathrm{M}}[\phi_\pm,\mathbf{A}_\pm]}\notag\\
&=\int \mathcal{D}[\mathbf{A}_+,\mathbf{A}_-]e^{iS[\mathbf{A}_+,\mathbf{A}_-]}\,,
\end{align}
where $S_{\mathrm{EM}}$ is the electromagnetic action and $S_{\mathrm{M}}$ contains the material sector and its coupling to electromagnetic fields. The second line defines the effective action $S$, obtained after integrating out the matter fields. 
$S$ is in general nonlocal in both space and time, and can be expanded in powers of the electromagnetic field with kernels of arbitrarily high order.
These kernels are the effective electromagnetic vertices generated by the material response.

After the Keldysh rotation of variables to the `classical' field   $\mathbf{A}_{\mathrm{cl}}=(\mathbf{A}_+ + \mathbf{A}_-)/2$ and `quantum' field   $\mathbf{A}_{\mathrm{q}}=\mathbf{A}_+ - \mathbf{A}_-$, the effective action can be expanded as:
\begin{align}
S=&S_0+S_{\mathrm{NL}}\notag
\\=&\sum_{\omega,\mathbf{k}}\mathbf{A}_{\mathrm{q}}(-\omega,-\mathbf{k})\Bigl[
\mathcal{D}_{\mathrm{vac},R}^{-1}(\omega,\mathbf{k})-\Pi_R(\omega,\mathbf{k})\Bigr]\mathbf{A}_{\mathrm{cl}}(\omega,\mathbf{k})
\notag\\
&+\sum_{\omega,\mathbf{k}}\mathbf{A}_{\mathrm{q}}(-\omega,-\mathbf{k})\Pi_K(\omega,\mathbf{k})\mathbf{A}_{\mathrm{q}}(\omega,\mathbf{k})+S_{\mathrm{NL}},
\label{eq:SeffKeldysh}
\end{align}
where $\mathcal{D}_{\mathrm{vac},R}$ is the bare retarded photon propagator in vacuum, $\Pi_R$ is the retarded polarization kernel of the material, $\Pi_K$ is the Keldysh kernel that describes fluctuations/occupations, and $S_{\mathrm{NL}}$ contains the nonlinear vertices. 
The classical equation of motion can be obtained by varying the action with respect to the quantum component $\mathbf{A}_{\mathrm{q}}$. 
The retarded linear electromagnetic response is controlled by the kernel appearing in the bilinear term $\mathbf{A}_{\mathrm{q}} \mathbf{A}_{\mathrm{cl}}$. 
It is convenient to define the bare retarded propagator in the vector-potential basis as:
\begin{align}
\mathcal{D}_{0,R}(\omega,\mathbf{k})&=
-i\langle \mathbf{A}_{\mathrm{cl}}(t,\mathbf{k}) \mathbf{A}_{\mathrm{q}}(0,\mathbf{k})  \rangle_0
\big|_{\omega}
\notag\\
&
=\frac{4\pi c^2}{\omega^2\!\left[1+4\pi\chi(\omega,\mathbf{k})\right]-c^2k^2}
\notag\\
&
=-\frac{c^2}{\omega^2}G_0(\omega,\mathbf{k}),
\label{eq:GA0}
\end{align}
where we have neglected the indices for notational simplicity,
$\langle \rangle_0$ means path integral average over $S_0$,
and $G_0$ is the electric-field propagator introduced in \equa{eqn:G_0}. 
The $S_{\mathrm{NL}}$ contains multi-point response functions, whose retarded components are nonlinear optical susceptibilities~\cite{mukamel2008partially,hansen2012nonlinear,dorfman2016nonlinear}.

Under a strong periodic pump field $\mathbf{A}_{\mathrm{p}}(t)=\mathbf{A}_{\mathrm{p}}e^{-i\omega_{\mathrm{p}} t}+\mathbf{A}_{\mathrm{p}} e^{i\omega_{\mathrm{p}} t}$, one replaces the classical component $\mathbf{A}_{\mathrm{cl}}(t)$ in Eq.~(\ref{eq:SeffKeldysh}) by
$
 \mathbf{A}_{\mathrm{p}}(t)+\mathbf{A}_{\mathrm{cl}}(t),
$
where $\mathbf{A}_{\mathrm{cl}}$ is  now viewed as a weak fluctuation. For example, a quartic vertex of the form $\mathbf{A}_{\mathrm q}(\mathbf{A}_{\mathrm{cl}})^3$ is now expanded as
$
\mathbf{A}_{\mathrm q}\mathbf{A}_{\mathrm p}^3+3\mathbf{A}_{\mathrm q}\mathbf{A}_{\mathrm{cl}}\mathbf{A}_{\mathrm p}^2+3\mathbf{A}_{\mathrm q}\mathbf{A}_{\mathrm{cl}}^2\mathbf{A}_{\mathrm p}+\mathbf{A}_{\mathrm q}\mathbf{A}_{\mathrm{cl}}^3
$.
The full retarded polariton propagator is 
\begin{align}
\mathcal{D}_{R}(t_1, t_2; \mathbf{k})=
-i\langle \mathbf{A}_{\mathrm{cl}}(t_1,\mathbf{k}) \mathbf{A}_{\mathrm{q}}(t_2,\mathbf{k})  \rangle
\end{align}
where the $\langle \cdots \rangle = \int \mathcal{D}[\mathbf{A}_{\mathrm{cl}},\mathbf{A}_{\mathrm{q}}] (\cdots)e^{i S}$ means the path integral average over the full action $S$.
The perturbative expansion of $\mathcal{D}_{R}$ in the pump field and in the 
polariton-polariton interactions in $S_{\mathrm{NL}}$ is performed in the standard way of expanding the exponential in these terms.
This process is well organized using diagrams made of free propagators and interaction vertices and the pump lines.
Note that $\mathcal{D}_{R}$ is a function of two times, meaning that its Fourier representation is a function of two frequencies:
$\mathcal{D}_{R}(t_1, t_2; \mathbf{k})= \sum_{\omega_1, \omega_2} \mathcal{D}_{R}(\omega_1, \omega_2; \mathbf{k}) e^{i(\omega_1 t_1- \omega_2 t_2)}$.
For the periodically driven problem, the difference between these two frequencies is restricted to integer multiples of the pump frequency $\omega_{\text{p}}$.
In the main text, we focused on the frequency-conserving component of the propagator: $\mathcal{D}_{R}(\omega, \mathbf{k}) \equiv \mathcal{D}_{R}(\omega, \omega; \mathbf{k})$, whose poles already give  the relevant Floquet polaritons.	
Its self-energy, the effective susceptibility, is also the most common observable in pump-probe experiments.
The diagrams in  \fig{fig:theoryFig} are the classical ones dominated by the pump.

\section{Green function and effective susceptibility for a uniaxial slab}\label{app:slabgreen}

\begin{figure}
	\includegraphics[width=\linewidth]{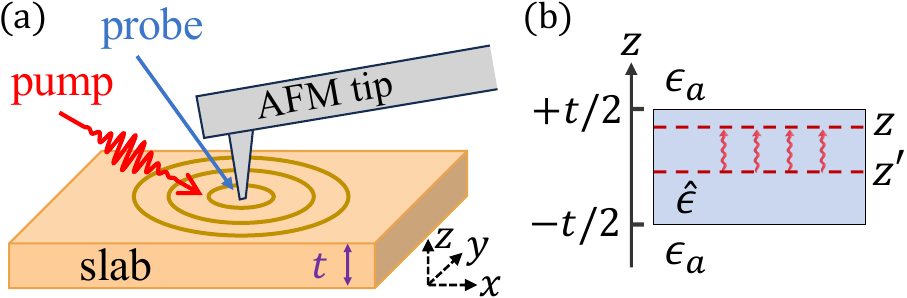}
	\caption{\label{fig:SIFig_slabG} 
(a) Schematic of the pump-probe experiment on a uniaxial slab of thickness $t$, embedded in a symmetric dielectric environment of permittivity $\epsilon_a$.
(b) Dielectric configuration used to derive the near-field Green function defined as  the field at $z$ generated by a  source polarization layer at $z'$.
}
\end{figure}

In this section, we show how to generalize the formalism in Sec.~\ref{sec:formalism}  to a finite-thickness sample with slab geometry. We consider a uniaxial film occupying $-t/2\leq z\leq t/2$ with the dielectric tensor
\begin{equation}
\hat{\varepsilon}(\omega)=\mathrm{diag}\!\left(\epsilon_{\perp}(\omega),\epsilon_{\perp}(\omega),\epsilon_z(\omega)\right),
\end{equation}
embedded in a symmetric environment of dielectric constant $\epsilon_a$, as shown in Fig.~\ref{fig:SIFig_slabG}. We consider the near-field limit so that one may write $\mathbf{E}=-\nabla\phi$, which is a good approximation for most of the polaritons in near field experiments. 
For a source polarization $\mathbf{P}(\omega,\mathbf{q},z)=P_x(\omega,\mathbf{q},z)\hat{e}_x$
where the in-plane momentum is $\mathbf{q}=q\hat{e}_x$,
the scalar potential inside the slab satisfies the Poisson equation:
\begin{equation}
\epsilon_z(\omega)\left(\partial_z^2-\kappa^2\right)\phi(\omega,\mathbf{q},z)=4\pi i q\, P_x(\omega,\mathbf{q},z),
\label{eq:slabphi}
\end{equation}
with
$
\kappa(\omega,q)=q\sqrt{\epsilon_{\perp}(\omega)/\epsilon_z(\omega)}.
$
Outside the slab, the potential satisfies $(\partial_z^2-q^2)\phi_{\rm out}=0$, whose decaying solutions are $\phi_{\rm out}^{(+)}(\mathbf{r})=A_+e^{iqx-q(z-t/2)}$ for $z>t/2$ and $\phi_{\rm out}^{(-)}(\mathbf{r})=A_-e^{iqx+q(z+t/2)}$ for $z<-t/2$. Continuity of $\phi$ and of the normal displacement field $D_z=-\epsilon_z \partial_z\phi$ across $z=\pm t/2$ then yields $A_+=\phi(z=t/2)$, $A_-=\phi(z=-t/2)$, and hence the boundary conditions:
\begin{subequations}\label{eq:boundaryconditions}
\begin{align}
\epsilon_z(\omega)\,\partial_z \phi\big|_{z=+t/2}&=-\epsilon_a q\,\phi(+t/2),\\
\epsilon_z(\omega)\,\partial_z \phi\big|_{z=-t/2}&=\epsilon_a q\,\phi(-t/2).
\end{align}
\end{subequations}
We define the slab Green function $g_q(z,z';\omega)$ by
\begin{equation}
\epsilon_z(\omega)\left(\partial_z^2-\kappa^2\right)g_q(z,z';\omega)
=
4\pi \delta(z-z')
\label{eq:slabgdef}
\end{equation}
together with the same boundary conditions in Eq.~(\ref{eq:boundaryconditions}). It is convenient to introduce the single-interface near-field reflection coefficient
\begin{equation}
r(\omega,q)=\frac{\epsilon_z(\omega)\kappa(\omega,q)-\epsilon_a q}
{\epsilon_z(\omega)\kappa(\omega,q)+\epsilon_a q}
\end{equation}
and the two solutions of the corresponding homogeneous equation:
\begin{subequations}\label{eq:homo_solutions}
\begin{align}
u_+(z)&=e^{\kappa(t/2-z)}+re^{-\kappa(t/2-z)},\\
u_-(z)&=e^{\kappa(t/2+z)}+re^{-\kappa(t/2+z)}.
\end{align}
\end{subequations}
By construction, $u_+(z)$ and $u_-(z)$ satisfy the boundary conditions at $z=+t/2$ and $z=-t/2$, respectively.
For $z\neq z'$ and $|z|\leq t/2$, the Green function must be constructed from these two solutions, up to a constant factor. Accordingly, we take
\begin{equation}
g_q(z,z';\omega)=
\begin{cases}
C\,u_-(z)\,u_+(z'), & -t/2\leq z<z',\\[4pt]
C\,u_-(z')\,u_+(z), & z'<z\leq t/2.
\end{cases}
\label{eq:slabG_guess}
\end{equation}
The remaining constant $C$ is fixed by the boundary conditions at $z=z'$. Since the source in Eq.~(\ref{eq:slabgdef}) is a delta function, $g_q(z,z';\omega)$ is continuous at $z=z'$, while its derivative obeys
\begin{equation}
\epsilon_z(\omega)\left[\partial_z g_q(z,z';\omega)\right]_{z=z'+0^-}^{z=z'+0^+}=4\pi.
\label{eq:slabgjump}
\end{equation}
Using Eq.~(\ref{eq:slabG_guess}), the Green function can be derived as
\begin{equation}
g_q(|z|\leq t/2,z';\omega)=-\frac{2\pi}{\epsilon_z(\omega)\kappa(\omega,q)}\frac{u_-(z_<)\,u_+(z_>)}{e^{\kappa t}\left(1-r^2 e^{-2\kappa t}\right)},
\label{eq:slabg}
\end{equation}
where $z_<\equiv \min(z,z')$ and $z_>\equiv \max(z,z')$. The poles of this Green function are determined by
\begin{equation}
1-r^2 e^{-2\kappa t}=0,
\end{equation}
which gives all the  polariton modes. When $\mathrm{Re}\,\epsilon_{\perp}$ and $\mathrm{Re}\,\epsilon_z$ have opposite signs, $\kappa$ becomes predominantly imaginary, so that the fields are propagating inside the slab while being evanescent outside. In this regime, the poles correspond to the hyperbolic polariton modes.

The in-plane electric field inside the slab is expressed as
\begin{equation}
E_x(\omega,\mathbf{q},z)=q^2\int_{-\frac{t}{2}}^{\frac{t}{2}} dz'\,g_q(z,z';\omega)P_x(\omega,\mathbf{q},z').
\end{equation}
Accordingly, the near-field propagator in the construction of Sec.~\ref{sec:formalism} is
\begin{equation}
G^{\mathrm{slab}}_{0}(\omega,\mathbf{q};z,z')=q^2g_q(z,z';\omega).
\label{eq:slabGxx}
\end{equation}
The effective susceptibility must then be generalized from a scalar quantity to a nonlocal object in $z$. 
Allowing the pump field to vary in $z$ direction, the slab generalization of Eq.~(\ref{eq:chieff}) in the near-field limit is obtained as
\begin{align}\label{eq:chieff_SI}
& \chi_{\mathrm{eff}}(\omega,\mathbf{q};z) = \sum_i \tilde{\chi}^{(i+1)}_{n=0}(\omega,\mathbf{q};z)E_{\mathrm p}^{i}(z) \notag\\
&+ \sum_{ij,n \neq 0}\int_{-\frac{t}{2}}^{\frac{t}{2}} dz'\tilde{\chi}^{(j+1)}_{-n}(\omega+n\omega_{\mathrm p},\mathbf{q};z)E_{\mathrm p}^{j}(z)\times\notag\\
&G^{\mathrm{slab}}_{0}(\omega+n\omega_{\mathrm p},\mathbf{q};z,z')\tilde{\chi}^{(i+1)}_{n}(\omega,\mathbf{q};z')E_{\mathrm p}^{i}(z').
\end{align}


Since the near-field propagator in Eq.~(\ref{eq:slabGxx}) inherits the pole structure of Eq.~(\ref{eq:slabg}), the basic Floquet-polariton mechanism remains the same as that discussed in Sec.~\ref{sec:formalism}. In particular, when the shifted frequency $\omega+n\omega_{\mathrm p}$ approaches a slab polariton mode, the resonant enhancement of the slab propagator strongly amplifies the corresponding contribution to $\chi_{\mathrm{eff}}$. In this sense, the  slab geometry only modifies the explicit form of the Green function.

\subsection{Two-dimensional limit}
\label{app:2Dlimit}
The two-dimensional near-field propagator in Eq.~(\ref{eq:G_0nearfield}) can be obtained from the slab result by taking the thin-film limit $t\to0$ while keeping the 2D sheet susceptibility fixed as
$
\chi_{\mathrm{2D}}(\omega,\mathbf q) = t\chi_{\mathrm{3D}}(\omega,\mathbf q)
$.
Here we take the two-dimensional layer to carry only an in-plane sheet
polarization and neglect its out-of-plane sheet response. Accordingly, we have $\epsilon_z\rightarrow \epsilon_a$ and
\begin{equation}
\epsilon_{\perp}(\omega,\mathbf q) = \epsilon_a+\frac{4\pi\chi_{\mathrm{2D}}(\omega,\mathbf q)}{t}\to \frac{4\pi\chi_{\mathrm{2D}}(\omega,\mathbf q)}{t}.
\label{eq:thinfilm_eps}
\end{equation}
so that
$
\kappa t \to q \sqrt{4\pi t \chi_{\mathrm{2D}}(\omega,\mathbf q)/\epsilon_a} 
$
and 
$
q/\kappa\to 0
$.
The single-interface near-field reflection coefficient becomes
\begin{equation}
r=\frac{\epsilon_z\kappa-\epsilon_a q}{\epsilon_z\kappa+\epsilon_a q}\approx 1-\frac{2q}{\kappa},
\end{equation}
and the denominator in Eq.~(\ref{eq:slabg}) that determines
the slab-polariton poles reduces to
\begin{align}
1-r^2e^{-2\kappa t}\approx\frac{4q}{\kappa}+2\kappa t = \frac{4q}{\kappa}\left(1+\frac{2\pi q\chi_{\mathrm{2D}}}{\epsilon_a}\right)
\label{eq:thinfilm_denominator}
\end{align}
 in the 2D limit.
Since $\kappa t\to0$, the exponential factors in Eq.~(\ref{eq:slabg}) approach unity. In particular, $u_+(0)u_-(0)\to4$, where we have made use of $r\to1$. Combining the results above, the slab Green function in Eq.~(\ref{eq:slabGxx}) reduces to
\begin{align}
G^{\mathrm{2D,nf}}_0(\omega,\mathbf q) 
&= \lim_{t\to0}q^2g_q(0,0;\omega) 
\notag\\
&= -\frac{2\pi q}{\epsilon_a+2\pi q\chi_{\mathrm{2D}}(\omega,\mathbf q)}.
\label{eq:G2Dnf_general}
\end{align}
For a two-dimensional layer embedded in vacuum, $\epsilon_a=1$, Eq.~(\ref{eq:G2Dnf_general}) reduces to Eq.~(\ref{eq:G_0nearfield}) in the main text.

The thin-film limit $t\to0$ also removes the explicit dependence on the slab coordinate $z$ in Eq.~(\ref{eq:chieff_SI}), as the pump and probe fields become uniform across the slab in the 2D limit.
We define the sheet nonlinear susceptibilities by integrating the corresponding three-dimensional susceptibilities across the slab thickness:
\begin{equation}
\tilde{\chi}^{(i+1)}_{n,\mathrm{2D}}(\omega,\mathbf q) = \int_{-t/2}^{t/2} dz\,
\tilde{\chi}^{(i+1)}_{n,\mathrm{3D}}(\omega,\mathbf q;z),
\label{eq:chi2D_def}
\end{equation}
and the sheet polarization is
\begin{equation}
P_x^{\mathrm{2D}}(\omega,\mathbf q) = \int_{-t/2}^{t/2} dz\,P_x(\omega,\mathbf q,z).
\end{equation}
Therefore, the expression for the slab susceptibility in Eq.~(\ref{eq:chieff_SI}) reduces to the 2D version
\begin{align}
\chi^{\mathrm{2D}}_{\mathrm{eff}}(\omega,\mathbf q)
=&
\sum_i
\tilde{\chi}^{(i+1)}_{n=0,\mathrm{2D}}(\omega,\mathbf q)
E_{\mathrm p}^i+
\notag\\
&
\sum_{ij,n\neq 0}
\tilde{\chi}^{(j+1)}_{-n,\mathrm{2D}}
(\omega+n\omega_{\mathrm p},\mathbf q)
E_{\mathrm p}^j \times
\notag\\
&
G^{\mathrm{2D,nf}}_0
(\omega+n\omega_{\mathrm p},\mathbf q)
\tilde{\chi}^{(i+1)}_{n,\mathrm{2D}}(\omega,\mathbf q)
E_{\mathrm p}^i .
\label{eq:chieff_2D_from_slab}
\end{align}
In this way, the explicit integrations over the slab coordinate are absorbed
into the sheet susceptibilities and the sheet Green function. Equation
(\ref{eq:chieff_2D_from_slab}) is the two-dimensional version of
Eq.~(\ref{eq:chieff}) used in the main text.

\section{Derivations for Floquet graphene plasmons}

\subsection{Near field reflection coefficient}
\label{appendix:Rp}

In this section, we show how the near-field reflection coefficient $R_{\mathrm{p}}$ in \fig{fig:GrPlasmon} is computed numerically. Starting from \equa{eqn:parametric_oscillator}, we expand the plasmonic field in the Floquet space as
\begin{equation}
	A_{\mathbf{q}}(t) =	\sum_{n=-N}^{N}	A_{\mathbf{q},n}e^{-i(\omega+2n\omega_{\mathrm{p}})t},
\end{equation}
where the cutoff $N=5$ is used in the calculation. Including damping by replacing $\Omega_n^2$ with $\Omega_n(\Omega_n+i\gamma)$ where $\Omega_n=\omega+2n\omega_{\mathrm{p}}$, \equa{eqn:parametric_oscillator} becomes
\begin{equation}
	\sum_{m=-N}^{N}\mathcal{M}_{nm}(\omega,q)A_{\mathbf{q},m}=S_n ,
\end{equation}
where $S_n$ is the external source in the $n$-th Floquet sector.
Here the Floquet matrix is 
\begin{equation}
	\mathcal{M}_{nm}(\omega,q)=M_n\delta_{nm}-V_q\left(\delta_{n,m+1}+\delta_{n,m-1}\right),
\end{equation}
where 
\begin{equation}
	M_n=2q\tilde{D}-\Omega_n(\Omega_n+i\gamma),
	\qquad  \tilde{D}=D\left(1-\frac{3\xi^2}{16}\right),
\end{equation}
and
\begin{equation}
	V_q=2\pi q c\alpha A_{\mathrm{p}}^2 =2qD\left(\frac{3\xi^2}{16}\right).\end{equation}
For a monochromatic near-field probe at frequency $\omega$, the source is applied to the $n=0$ sector, and the measured response is also the $n=0$ component.

The effective kernel in the physical probe sector is obtained by integrating out all sideband sectors:
\begin{equation}
	M_{\mathrm{eff}}(\omega,q)=\left[\left(\mathcal{M}^{-1}\right)_{00}\right]^{-1}.
\end{equation}
In practice, this quantity is evaluated by the continued-fraction recursion of the tridiagonal Floquet matrix up to the cutoff $N=5$.
The effective Floquet dielectric function is then defined as
\begin{equation}
	\epsilon_{\mathrm{2D}}^{\mathrm{F}}(\omega,q)=-\frac{M_{\mathrm{eff}}(\omega,q)}{\omega(\omega+i\gamma)}
\end{equation}
which also defines a pump-dressed effective susceptibility $\chi_{\mathrm{eff}}^{\mathrm{F}}(\omega,q)$ through
$
	\epsilon_{\mathrm{2D}}^{\mathrm{F}}(\omega,q)=1+2\pi q\chi_{\mathrm{eff}}^{\mathrm{F}}(\omega,q)
$.
Without pump, the dielectric reduces to
\begin{equation}
	\epsilon_{\mathrm{2D}}(\omega,q)=1-\frac{2qD}{\omega(\omega+i\gamma)}.
\end{equation}
This construction directly connects to the effective-susceptibility  in the main text. Keeping only the $n=\pm1$ sidebands  gives
\begin{equation}
	M_{\mathrm{eff}}(\omega,q)\simeq M_0-\frac{V_q^2}{M_{1}}-\frac{V_q^2}{M_{-1}},
\end{equation}
and hence
\begin{align}
	&\chi_{\mathrm{eff}}^{\mathrm{F}}(\omega,q)=-\frac{\tilde{D}}{\pi\omega(\omega+i\gamma)}\notag\\&+\frac{V_q^2}{2\pi q\,\omega(\omega+i\gamma)}\bigg[\frac{1}{2q\tilde{D}-(\omega+2\omega_{\mathrm{p}})(\omega+2\omega_{\mathrm{p}}+i\gamma)}\notag\\&\qquad\qquad\qquad\qquad
	+(\omega_\mathrm{p}\to-\omega_\mathrm{p})\bigg].
\end{align}
which reproduces \equa{eq:grchieff} after including damping.

Finally, the near-field reflection coefficient is calculated numerically as
\begin{equation}
	R_{\mathrm{p}}(\omega,q)=1-\frac{1}{\epsilon_{\mathrm{2D}}^{\mathrm{F}}(\omega,q)} 
\end{equation}
which yields the color-scale plots in \fig{fig:GrPlasmon}(c).
The eigenfrequencies of the Floquet-plasmons (black curves) there are obtained from the homogeneous Floquet equation
\begin{equation}
	\det\mathcal{M}(\omega,q)=0.
\end{equation}

\begin{figure}[t]
	\includegraphics[width=\linewidth]{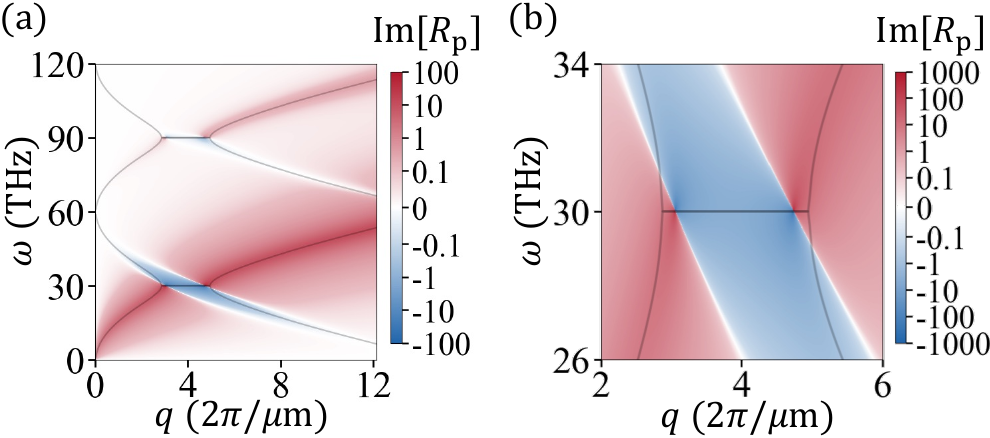}
	\caption{\label{fig:GrPlasmonLargeGamma}
		(a)  Near-field reflection coefficient $R_{\text{p}}$ of Floquet graphene plasmons at a larger damping rate $\gamma=5\unit{THz}$, but with
		all other parameters the same as those in \fig{fig:GrPlasmon} at the  pump strength $E_{\mathrm{p}}=600\unit{kV/cm}$.
		The color scale shows  $\mathrm{Im}[R_{\text{p}}]$ while the black curves denote the real parts of the complex Floquet-plasmon eigenfrequencies.
		(b) Zoom-in plot of (a) near the band crossing region.
		Divergence of $\mathrm{Im}[R_{\text{p}}]$ appears at the two threshold  points of parametric instability, which are at different locations from	 the  exceptional points, i.e., points where the black curves start to merge.
	}
\end{figure}

To clarify the relation between the real-frequency reflection signal and the complex Floquet eigenmode spectrum, we show in \fig{fig:GrPlasmonLargeGamma} the near-field reflection coefficient for a larger plasmonic damping rate. All parameters are chosen to be the same as those in \fig{fig:GrPlasmon}, with the nonzero pump field strength fixed at $E_{\mathrm{p}}=600\unit{kV/cm}$, except that the damping rate is increased to $\gamma=5\unit{THz}$.
Near the crossing between the original plasmon branch and its Floquet replica, the two relevant modes can be approximately described by \equa{eq:floquetplasmon_freq}. The exceptional points occur at $|Q|=\kappa$, where the two complex eigenfrequencies coalesce.  Inside the momentum-gap region, $|Q|<\kappa$, \equa{eq:floquetplasmon_freq} becomes
\begin{equation}
	\omega_q=\omega_{\mathrm{p}}-i\frac{\gamma}{2}\pm i\omega_{\mathrm{p}}\sqrt{\kappa^2-Q^2}.
\end{equation}
However, the near-field reflection coefficient is a real-frequency response function.
The strongest real-frequency response appears when one of the poles is closest to the real-frequency axis.
This happens at the threshold of parametric instability:
\begin{equation}
	Q^2	=\kappa^2-\frac{1}{4}\left(\frac{\gamma}{\omega_{\mathrm{p}}}\right)^2
\end{equation}
where one of the two modes has absolutely zero damping rate.
Note that this condition does not coincide with that ($Q^2=\kappa^2$) of the exceptional-point.
Their  difference is difficult to see when the intrinsic damping is small as these points are close to each other on the frequency-momentum plane.
For the larger damping rate $\gamma=5\unit{THz}$, the separation becomes clearly visible.
As shown in \fig{fig:GrPlasmonLargeGamma}(b), two singularities of $\mathrm{Im}[R_{\mathrm{p}}]$ appear at the momenta different from the two exceptional points where the real parts of the two Floquet eigenfrequencies start to merge.

\subsection{Effects of circularly polarized pump}
\label{appendix:circular}

For completeness, we briefly discuss the effects of a circularly polarized pump on graphene plasmons. To keep the net pump strength consistent with the linearly polarized case in the main text, we define the circularly polarized pump field as
\begin{equation}
	\mathbf{A}_{\mathrm{p}}(t)=\frac{A_{\mathrm{p}}}{\sqrt{2}}\left[\cos(\omega_{\mathrm{p}}t)\hat{\mathbf{x}}+\eta\sin(\omega_{\mathrm{p}}t)\hat{\mathbf{y}}\right],
	\label{eq:circular_Ap}
\end{equation}
where $\eta=\pm1$ labels the pump chirality. We therefore define the same dimensionless pump strength as $\xi=eE_{\mathrm{p}}/(\hbar k_{\mathrm{F}}\omega_{\mathrm{p}})$.

From \equa{eq:grapheneL}, the intraband current gives
\begin{align}
	j_i(t)=&-\frac{D}{\pi c}A_i(t) +
	\notag\\
	&\frac{3D^{(3)}}{c^3}\left[\mathbf{A}_{\mathrm{p}}^2(t)\delta_{ij}+2A_{\mathrm{p},i}(t)A_{\mathrm{p},j}(t)\right]A_j(t).
	\label{eq:circular_jA_tensor}
\end{align}
as a linear response to the weak probe field.
For a longitudinal plasmon, $\mathbf{A}_{\mathbf{q}}=A_{\mathbf{q}}\hat{\mathbf{q}}$, projecting \equa{eq:circular_jA_tensor} onto $\hat{\mathbf{q}}$ gives
\begin{equation}
	j_{\mathbf{q}}(t)=-\frac{D}{\pi c}A_{\mathbf{q}}(t)+\frac{3D^{(3)}}{c^3}\left[\mathbf{A}_{\mathrm{p}}^2(t)+2\left(\mathbf{A}_{\mathrm{p}}(t)\cdot\hat{\mathbf{q}}\right)^2\right]A_{\mathbf{q}}(t),
	\label{eq:circular_jq_Aq_general}
\end{equation}
where $\hat{\mathbf{q}}=(\cos\theta_{\mathbf{q}},\sin\theta_{\mathbf{q}})$. For the circularly polarized pump in \equa{eq:circular_Ap}, one has
\begin{equation}
	\mathbf{A}_{\mathrm{p}}^2(t)=\frac{A_{\mathrm{p}}^2}{2},
	\qquad
	\mathbf{A}_{\mathrm{p}}(t)\cdot\hat{\mathbf{q}}=\frac{A_{\mathrm{p}}}{\sqrt{2}}\cos(\omega_{\mathrm{p}}t-\eta\theta_{\mathbf{q}}).
\end{equation}
Substituting these relations into \equa{eq:circular_jq_Aq_general}, one obtains
\begin{align}
	j_{\mathbf{q}}(t)=&-\frac{D}{\pi c}A_{\mathbf{q}}(t)\notag\\
	&+\frac{3D^{(3)}A_{\mathrm{p}}^2}{c^3}\left[1+\frac{1}{2}\cos 2(\omega_{\mathrm{p}}t-\eta\theta_{\mathbf{q}})\right]A_{\mathbf{q}}(t).
	\label{eq:circular_jq_Aq}
\end{align}
The zero-Fourier component of the pump-induced term renormalizes the bare Drude response isotropically, giving
\begin{equation}
	\tilde{D}_{\mathrm{circ}} = D\left(1-\frac{\xi^2}{8}\right).
	\label{eq:Dtilde_circular}
\end{equation}
The  time-periodic component reads
\begin{equation}
	\cos 2(\omega_{\mathrm{p}}t-\eta\theta_{\mathbf{q}})=\frac{1}{2}e^{-2i(\omega_{\mathrm{p}}t-\eta\theta_{\mathbf{q}})}+\frac{1}{2}e^{2i(\omega_{\mathrm{p}}t-\eta\theta_{\mathbf{q}})},
\end{equation}
which couples Floquet components separated by $\pm2\omega_{\mathrm{p}}$ and carries the phase factor $e^{\mp2i\eta\theta_{\mathbf{q}}}$. These phases arise from projecting onto the longitudinal plasmon direction under the circularly polarized pump field.

Within a scalar description of the longitudinal plasmon response, the two Floquet vertices in the process $\omega\rightarrow \omega\pm2\omega_{\mathrm{p}}\rightarrow\omega$ carry complex-conjugate angular phase factors. Their product is therefore independent of $\theta_{\mathbf{q}}$, so the angular phase does not appear in the scalar effective susceptibility. Consequently, the Floquet plasmon dispersion in circularly pumped graphene remains isotropic at this level.

To see the tensor structure of the pump-dressed response, we return to \equa{eq:circular_jA_tensor}. For the circularly polarized pump, its pump-induced tensor components are
\begin{align}
	&\mathbf{A}_{\mathrm{p}}^2(t)\delta_{ij}+2A_{\mathrm{p},i}(t)A_{\mathrm{p},j}(t)\notag\\
	=&A_{\mathrm{p}}^2\delta_{ij}+\frac{A_{\mathrm{p}}^2}{2}
	\begin{pmatrix}
		\cos2\omega_{\mathrm{p}}t & \eta\sin2\omega_{\mathrm{p}}t\\
		\eta\sin2\omega_{\mathrm{p}}t & -\cos2\omega_{\mathrm{p}}t
	\end{pmatrix}_{ij}
	\notag\\
	=&A_{\mathrm{p}}^2\delta_{ij}+\frac{A_{\mathrm{p}}^2}{4}\left[K_{+}e^{-2i\omega_{\mathrm{p}}t}+K_{-}e^{2i\omega_{\mathrm{p}}t}\right]_{ij},
	\label{eq:circular_tensor_decomp}
\end{align}
where
\begin{equation}
	K_{+}=
	\begin{pmatrix}
		1 & i\eta\\
		i\eta & -1
	\end{pmatrix},\qquad
	K_{-}=
	\begin{pmatrix}
		1 & -i\eta\\
		-i\eta & -1
	\end{pmatrix}.
\end{equation}
The first term in \equa{eq:circular_tensor_decomp} gives the isotropic renormalization of the Drude weight in \equa{eq:Dtilde_circular}. The last term gives the Floquet vertices connecting frequencies separated by $2\omega_{\mathrm{p}}$.
For the process $\omega\rightarrow\omega+2\omega_{\mathrm{p}}\rightarrow\omega$, the intermediate plasmon is longitudinal, with projector $\mathcal{P}^{\mathrm{L}}_{ij}=\hat{\mathbf{q}}_i\hat{\mathbf{q}}_j$. Hence the tensor factor is
\begin{equation}
	K_{-}\mathcal{P}^{\mathrm{L}}K_{+}=\delta_{ij}+i\eta\epsilon_{ij}.
\end{equation}
Similarly, the process $\omega\rightarrow\omega-2\omega_{\mathrm{p}}\rightarrow\omega$ gives
\begin{equation}
	K_{+}\mathcal{P}^{\mathrm{L}}K_{-}=\delta_{ij}-i\eta\epsilon_{ij}.
\end{equation}
Therefore, although the angular phases cancel in the scalar longitudinal dispersion, the full in-plane response retains a chirality-dependent antisymmetric tensor structure.

Keeping the same order of Floquet processes as in \equa{eq:grchieff}, the effective susceptibility can be written as
\begin{equation}
	\chi^{\mathrm{circ}}_{\mathrm{eff},ij}=\chi_{\mathrm{S}}(\omega,q)\delta_{ij}+i\chi_{\mathrm{H}}(\omega,q)\epsilon_{ij},
	\label{eq:circular_decomposition}
\end{equation}
with
\begin{align}
	\chi_{\mathrm{S}}(\omega,q)=&
	-\frac{\tilde{D}_{\mathrm{circ}}}{\pi\omega(\omega+i\gamma)}\notag\\
	&+\frac{D}{\pi}\frac{2qD}{\omega^2}\left(\frac{\xi^2}{16}\right)^2
	\left[\frac{1}{\mathcal{D}_{+}(\omega,q)}+\frac{1}{\mathcal{D}_{-}(\omega,q)}\right],
	\label{eq:circular_chi_S}
	\\
	\chi_{\mathrm{H}}(\omega,q)=&\eta\frac{D}{\pi}\frac{2qD}{\omega^2}\left(\frac{\xi^2}{16}\right)^2\left[\frac{1}{\mathcal{D}_{+}(\omega,q)}-\frac{1}{\mathcal{D}_{-}(\omega,q)}\right],
	\label{eq:circular_chi_H}
\end{align}
where
\begin{equation}
	\mathcal{D}_{\pm}(\omega,q)=2q\tilde{D}_{\mathrm{circ}}-(\omega\pm2\omega_{\mathrm{p}})(\omega\pm2\omega_{\mathrm{p}}+i\gamma).
	\label{eq:circular_Dpm}
\end{equation}
Here $\chi_{\mathrm{S}}$ is the scalar symmetric part, while $\chi_{\mathrm{H}}$ is a  Hall response that changes sign under reversal of the pump chirality.

For the purely longitudinal plasmon in the bulk of the 2D sheet, one has
$
	\hat{\mathbf{q}}_i\epsilon_{ij}\hat{\mathbf{q}}_j=0,
$
so the plasmon condition only involves $\chi_{\mathrm{S}}$:
\begin{equation}
	\epsilon_{\mathrm{2D}}=1+2\pi q\,\hat{\mathbf{q}}_i\chi^{\mathrm{circ}}_{\mathrm{eff},ij}\hat{\mathbf{q}}_j=1+2\pi q\chi_{\mathrm{S}} = 0.
\end{equation}
Physically, the Hall current induced by the longitudinal electric field in the plasmonc oscillation is perpendicular to the wave vector, so that it does not lead to charge accumulation and thus does not affect the equation of motion.
Therefore, the pump-induced Hall response does not affect the bulk plasmon dispersion, similar to the case of 2D electron gas in a static magnetic field.
However, it may become important at the boundary of the sample.

\subsection{Effects from  second order nonlinearity}
\label{appendix:chi2}
Assuming both the plasmonic momentum $\mathbf{q}$ (and thus its electric field) and the uniform pump field $\mathbf{A}_{\text{p}}=A_{\text{p}}\cos(\omega_{\text{p}} t)\hat{x}$ are  along $\hat{x}$,  the second order nonlinear current  of  a 2D electron fluid is~\cite{sun2018universal}
\begin{align}\label{eq:j2}
	j_{\mathbf{q}}^{(2)}(\omega\pm\omega_{\text{p}}) = -qD^{(2)}A_{\text{p}}A_{\mathbf{q}}
	\left(\frac{1}{\omega}+\frac{1}{\omega\pm\omega_{\text{p}}}
	\right)
	\,.
\end{align}
in the hydrodynamic regime.
As a result, the self-energy in \fig{fig:theoryFig}(c) reads
\begin{align}
	\chi^{(c)}(\omega,\mathbf{q})= &
	\frac{2\pi q(qcD^{(2)}A_{\text{p}}/\omega)^2}{2q\tilde{D}-(\omega+\omega_{\text{p}})^2}
	\left(\frac{1}{\omega}+\frac{1}{\omega+\omega_{\text{p}}}
	\right)^2
	\notag\\
	& + (\omega_{\text{p}} \rightarrow -\omega_{\text{p}}).
	\label{eq:grchiefftotal}
\end{align}
Floquet polariton branches shifted by $\omega_{\text{p}}$ emerge due to this effective susceptibility. 
Specifically, the second term of \equa{eq:grchiefftotal} indicates a band crossing point  at $\omega=\omega_{\text{p}}/2$ between the original plasmon band and  the $n=1$ replica.
However, their hybridization is absent because the strength $\sim \left(\frac{1}{\omega}+\frac{1}{\omega-\omega_{\text{p}}}
\right)^2$ of this term vanishes at this frequency. 
Therefore, a careful analysis of the second order susceptibility reveals that it contributes only an additional set of Floquet polariton branches shifted by $N \omega_{\text{p}}$, but without inducing any nontrivial effects. 
The second order nonlinear response in the high frequency kinetic regime~\cite{sun2018universal} gives the same result.

The absence of parametric instability can be understood from inversion symmetry. Parametric down conversion requires the normally incident pump photon to convert	 into two plasmons with opposite momenta in this system. 
Such a coupling term $\mathbf{A}_{\text{p}} A_{\mathbf{q}} A_{-\mathbf{q}}$ between the pump field and the two plasmonic fields is odd under inversion, which must therefore vanish~\cite{Sun_plasmon.2022}.
Furthermore, \equa{eq:j2} has the same structure as that of a Galilean invariant electron fluid, and must lead to the same effects as what it does for the latter.
From a Galilean invariant electron fluid, the plasmon dispersion should stay the same as the equilibrium one in the frame oscillating together with the fluid driven by the uniform pump.
There nontrivial effects such as parametric instability is obviously absent.
The Floquet plasmon dispersion can be obtained by transforming back to the lab frame, which only leads to trivial replicas of the original plasmonic bands.

\begin{widetext}
\section{Derivations for Floquet phonon polaritons in hBN}
\label{appendix:hBN}

\begin{figure}
	\includegraphics[width=0.7 \linewidth]{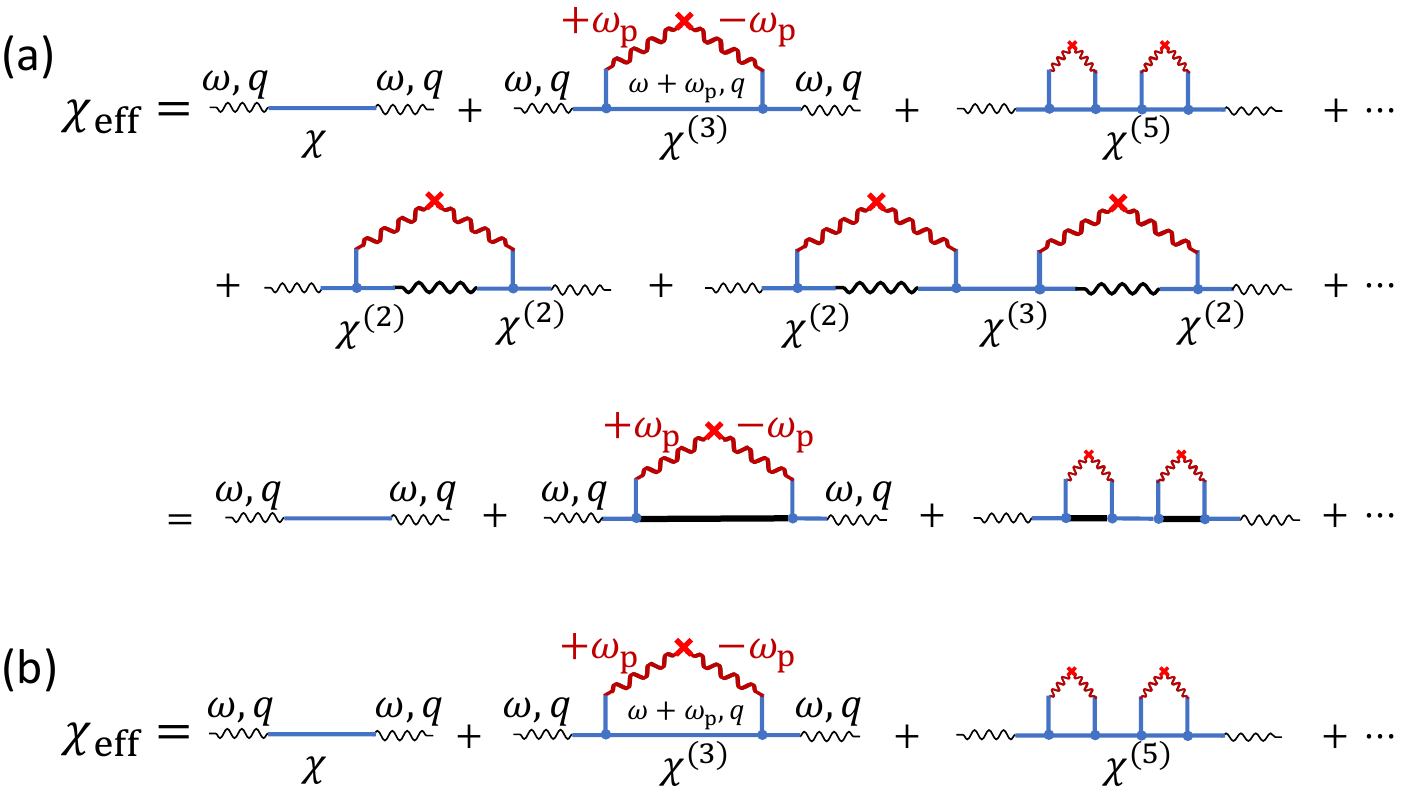}
	\caption{\label{fig:SIFig2} 
		(a) The diagrammatic representation of the effective susceptibility of laser-pumped monolayer hBN modeled by \equa{eq:hBNlagrangian}. 
		The blue solid lines are the bare phonon propagators:
		$G_{\text{T}}^{\text{phonon}}(\omega,\mathbf{q}) 
		= \frac{1}{m(\omega_{\text{TO}}^2 - \omega^2)}$.
		The red wavy line denotes the pump field, following \fig{fig:theoryFig}.
		The black thick wavy lines are polariton propagators in \equa{eq:phonon_polariton}.
		The black solid lines are polariton propagators defined in terms of the phononic response function:   $G_0^{\text{phonon}}(\omega,\mathbf{q}) 
		= \frac{1}{m(\omega_{\mathbf{q},L}^2 - \omega^2)}$.
		These diagrams are selected following principles: 
		1, They are 1PI diagrams that cannot be separated by cutting a photon propagator at frequency $\omega$, which is the definition of effective susceptibility; 
		2, They are at the classical level, or in other words, tree level if one views the pump field lines as external legs; 
		3, Only propagators with frequencies $\omega, \omega\pm \omega_{\text{p}}$ are involved, meaning that only the replica bands at order $n=0, \pm1$ are included.
		(b) Same as (a) but for bulk hBN, so that the diagrams with $\chi^{(2)}$ vertices connected to photons/polaritons all vanish because of cancellation between adjacent atomic layers of hBN.
	}
\end{figure}

For monolayer hBN, the effective susceptibility in \equa{eq:chieffhbn} contains the series of diagrams in \fig{fig:SIFig2}(a).
There only propagators with frequencies $\omega, \omega\pm \omega_{\text{p}}$ are involved, meaning that only the replica bands at order $n=0, \pm1$ are included, equivalent to solving the Floquet matrix with a cutoff of $n=\pm 1$.
Following the guidance of the diagrams, the series is summed as 
\begin{align}\label{eq:chihbn}
\chi_{\text{eff}}(\omega,\mathbf{q})
&=
\chi(\omega,\mathbf{q})
\frac{1}{1-
\left\{G_{\text{T}}^{\text{phonon}}(\omega,q)
\left[
E_{\text{p}} \lambda_0
G_{\text{T}}^{\text{phonon}}(\omega_{\text{p}},0)
\right]
6 \lambda_1
G_0^{\text{phonon}}(\omega+\omega_{\text{p}},q)
6 \lambda_1
\left[
E_{\text{p}} \lambda_0
G_{\text{T}}^{\text{phonon}}(-\omega_{\text{p}},0)
\right]
+ (\omega_{\text{p}} \rightarrow -\omega_{\text{p}} )
\right\}
}
\notag\\
&=
\frac{\lambda_0^2/m}
{\omega_{\text{TO}}^2 - \omega^2-
\left[
\frac{36}{m}
		E_{\text{p}}^2 \lambda_1^2  \lambda_0^2
	G_{\text{T}}^{\text{phonon}}(\omega_{\text{p}},0)
	G_0^{\text{phonon}}(\omega+\omega_{\text{p}},q)
	G_{\text{T}}^{\text{phonon}}(-\omega_{\text{p}},0)
+ (\omega_{\text{p}} \rightarrow -\omega_{\text{p}} )
\right]
}
\,.
\end{align}

For bulk hBN in the continuous limit, the diagrams with $\chi^{(2)}$ vertices connected to photons/polaritons  from \fig{fig:SIFig2}(a) all vanish because of cancellation between adjacent atomic layers of hBN.
The remaining nonzero diagrams are those involving odd-order optical nonlinearities only, as shown in \fig{fig:SIFig2}(b).
Compared to \equa{eq:chihbn}, this series only changes the 	full phonon propagator $G_0^{\text{phonon}}$ (that includes polaritonic effect) to the bare phonon propagator $G_{\text{T}}^{\text{phonon}}$:
\begin{align}\label{eq:chihbn_bulk}
	\chi_{\text{eff}}(\omega,\mathbf{q})
&=
\chi(\omega,\mathbf{q})
\frac{1}{1-
	\left\{G_{\text{T}}^{\text{phonon}}(\omega,q)
	\left[
	E_{\text{p}} \lambda_0
	G_{\text{T}}^{\text{phonon}}(\omega_{\text{p}},0)
	\right]
	6 \lambda_1
	G_{\text{T}}^{\text{phonon}}(\omega+\omega_{\text{p}},q)
	6 \lambda_1
	\left[
	E_{\text{p}} \lambda_0
	G_{\text{T}}^{\text{phonon}}(-\omega_{\text{p}},0)
	\right]
	+ (\omega_{\text{p}} \rightarrow -\omega_{\text{p}} )
	\right\}
}
	\notag\\
	&=
\frac{\lambda_0^2/m}
{\omega_{\text{TO}}^2 - \omega^2-
	\left[
	\frac{36}{m}
	E_{\text{p}}^2 \lambda_1^2  \lambda_0^2
	\frac{1}{m^2(\omega_{\text{TO}}^2 - \omega^2)^2}
	G_{\text{T}}^{\text{phonon}}(\omega+\omega_{\text{p}},q)
	+ (\omega_{\text{p}} \rightarrow -\omega_{\text{p}} )
	\right]
}	
	\,.
\end{align}

\end{widetext}

\begin{acknowledgments}
	This work is supported by Beijing Natural Science Foundation (Z240005),  the National Natural Science Foundation of China (Grants No. 12374291 and No. 12421004), the National Key Research and Development Program of China (Grants No. 2022YFA1204700 and 2022YFA1404704), 
    and the startup grant from Tsinghua University. 
    We thank Q. Xiong, Y. Xu, Z. Wang, E. Demler, M. Michael and J. Zhang for helpful discussions.
\end{acknowledgments}

\bibliography{reference}

\end{document}